\documentclass[12pt,fleqn]{article}

\usepackage{amsmath, amssymb,graphicx}

\renewcommand{\a}{\alpha}
\renewcommand{\b}{\beta}
  
\renewcommand{\d}{\delta}
  \newcommand{\s}{\sigma}
  \newcommand{\m}{\mu}

  \newcommand{\D}{\Delta}
  \newcommand{\G}{\Gamma}

  \newcommand{\N}{\mathcal{N}}

\newcommand{\Qsl}{Q\!\!\!\!\slash\,}
\newcommand{\Dsl}{D\!\!\!\!\slash\,}
\newcommand{\thetasl}{\theta\!\!\!\slash}

\def\pr{\partial}

\newcommand{\beq}{\begin{equation}}
\newcommand{\eeq}{\end{equation}}
\newcommand{\beqa}{\begin{eqnarray}}
\newcommand{\eeqa}{\end{eqnarray}}
\newcommand{\beqar}{\begin{eqnarray*}}
\newcommand{\eeqar}{\end{eqnarray*}}

\newcommand{\eps}{\epsilon}

\newcommand{\p}{\phi}

\newcommand{\labell}[1]{\label{#1}} %{\label{#1}\qquad_{#1}}%
\newcommand{\reef}[1]{(\ref{#1})}
\newcommand\prt{\partial}

\newcommand\Tr{{\rm Tr}}

\def\p1{\phantom{1}}
\def\IR{{\hbox{{\rm I}\kern-.2em\hbox{\rm R}}}}

\begin{document}

%%%%%%%%%%%%%%%%   TITLE    %%%%%%%%%%%%%%%%%%%%

\thispagestyle{empty}
\renewcommand{\thefootnote}{\fnsymbol{footnote}}

{\hfill \parbox{4cm}{
        HU-EP-02/32 \\
        MIT-CTP-3316 \\
}}

\bigskip\bigskip

\begin{center} \noindent \Large \bf
Intersecting D3-branes and Holography
\end{center}

\bigskip\bigskip\bigskip

\centerline{ \normalsize \bf Neil R. Constable$\,^{a}$, Johanna
Erdmenger$\,^b$,
Zachary Guralnik$\,^{b}$ and
Ingo Kirsch$\,^{b}$\footnote[1]{\noindent \tt
constabl@lns.mit.edu, jke@physik.hu-berlin.de,
zack@physik.hu-berlin.de,  ik@physik.hu-berlin.de} }

\bigskip
\bigskip\bigskip

\centerline{$^a$ \it
Center for Theoretical Physics and Laboratory for Nuclear Science    }
\centerline{ \it Massachusetts Institute of Technology}
\centerline{\it 77 Massachusetts Avenue}
\centerline{ \it Cambridge, {\rm MA}  02139, USA}
\bigskip
\centerline{$^b$ \it Institut f\"ur Physik}
\centerline{\it Humboldt-Universit\"at zu Berlin}
\centerline{\it Invalidenstra{\ss}e 110}
\centerline{\it D-10115 Berlin, Germany}
\bigskip\bigskip

\bigskip\bigskip

\renewcommand{\thefootnote}{\arabic{footnote}}

\centerline{\bf \small Abstract}
\medskip

{\small We study a defect conformal field theory describing
D3-branes intersecting over two space-time dimensions.
This theory admits an exact Lagrangian description
which includes both two- and four-dimensional
degrees of freedom, has $(4,4)$ supersymmetry and is invariant
under global conformal transformations. Both two- and
four-dimensional contributions to the action are conveniently
obtained in a two-dimensional $(2,2)$ superspace. In a suitable
limit, the theory has a dual description in terms of a probe
D3-brane wrapping an $AdS_3 \times S^1$ slice of $AdS_5\times
S^5$. We consider the AdS/CFT dictionary for this set-up. In
particular we find classical probe fluctuations corresponding to
the holomorphic curve $wy=c\alpha^{\prime}$. These fluctuations
are dual to defect fields containing massless two-dimensional
scalars which parameterize the classical Higgs branch, but do not
correspond to states in the Hilbert space of the CFT.  We also
identify probe fluctuations which are dual to BPS superconformal
primary operators and to their descendants. A non-renormalization
theorem is conjectured for the correlators of these operators, and
verified to order $g^2$.}

\newpage

%%%%%%%%%%%%%%%%%%%%%%%%%%%%%%%%%%%%%%%%%%%%%%%%%%%%%%%%%

%\bigskip

%\author{{\bf Neil Constable} \thanks{email: constabl@lns.mit.edu} \\ \\
%{\it Center for Theoretical Physics and Laboratory for Nuclear Science} \\
%{\it Massachusetts Institute of Technology} \\
%{\it Cambridge, MA 02139, USA} \\\\ \\
%{\bf Johanna Erdmenger, Zachary Guralnik and
%    Ingo Kirsch} \thanks{email: jke@physik.hu-berlin.de,
%    zack@physik.hu-berlin.de,
%    ik@physik.hu-berlin.de}\\ \\
%  {\it Institut f\"ur Physik}\\
%  {\it Humboldt-Universit\"at zu Berlin}\\
%  {\it Invalidenstra{\ss}e 110}\\
%  {\it D-10115 Berlin, Germany} }
%\date{}
%\maketitle
%\begin{abstract}
%\smallskip
%\smallskip
%\end{abstract}

%%%%%%%%%%%%%%%%%%%%%%%%%%%%%%%%%%%%%%%%%%%%%%%%%%%%%%%%%

\newpage
\section{Introduction and summary}
\setcounter{equation}{0}

The general problem of introducing a spatial defect into a
conformal field theory has been studied in several
contexts~\cite{Cardy,Osborn}.  Within string theory such defect
conformal field theories arise in various brane constructions.
They were first studied in this context as matrix model
descriptions of compactified NS5-branes \cite{Sethi} and more
generally as effective field theories describing various D-brane
intersections~\cite{GanorSethi, KapustinSethi}. More recently, an
extension of AdS/CFT duality~\cite{Juan} was conjectured in which
an $AdS_5\times S^5$ background is probed with a D5-brane wrapping
an $AdS_4 \times S^2$ submanifold. This configuration has been
conjectured to be dual to a four-dimensional conformal field
theory coupled to  a codimension one defect \cite{KR}.  This
defect conformal field theory describes the decoupling limit of
the D3-D5 intersection, and consists of the ${\cal N} = 4, d=4$
super Yang-Mills theory coupled to an ${\cal N}=4, d=3$
hypermultiplet localized at the defect. The open string modes with
both endpoints on the D5-brane decouple in the infrared.
Holographic duality can be viewed as acting twice: The ${\cal N}
=4, d=4$ super Yang-Mills degrees of freedom are dual to closed
strings in $AdS_5\times S^5$, while the defect hypermultiplet
degrees of freedom are dual to open strings with endpoints on the
probe D5-brane wrapping $AdS_4 \times S^2$. In \cite{DFO}, the
action of the model was written down, and the chiral primaries
localized on the defect were identified along with the dual
fluctuations on the $AdS_4$ brane. In \cite{EGK}, the action was
written compactly in an ${\cal N} =2, d=3$ superspace,  and field
theoretic arguments for quantum conformal invariance were given.
The supersymmetry of the $AdS_4 \times S^2$ embedding was
demonstrated in \cite{SkenderisTaylor}. Gravitational aspects of
this set-up were discussed in \cite{Porrati,Fayya,Liu}. The
Penrose limit of this background was studied in
\cite{SkenderisTaylor,LeePark}, wherein a map between defect
operators with large $R$-charge and open strings on a D3-brane in
a plane wave background was constructed. Moreover, two-dimensional
conformal field theories with a one-dimensional defect dual to
$AdS_2$ branes in $AdS_3$ have recently been studied in
\cite{Bachas,Schomerus}. In \cite{Katz, Katz2} spacetime filling
D7-branes were added to the $AdS_5/{\rm CFT}_4$ correspondence in order
to study flavors in supersymmetric extensions of QCD. Similarly the
supergravity solution for the D2/D6 intersection, dual to
2+1-dimensional Yang-Mills with flavor, was obtained in
\cite{Cherkis}. RG flows
related to defect conformal field theories were discussed in
\cite{Yama}. Finally, defect CFT's were discussed in connection
with the phenomenon of supertubes in~\cite{Mateos}.

In this paper we consider a defect conformal field theory which
describes the low energy dynamics of intersecting D3-branes.  This
system consists of a stack of D3-branes spanning the $0123$
directions and an orthogonal stack spanning the $0145$ directions
such that eight supercharges are preserved, realizing a
a $(4,4)$ supersymmetry on the
common $1+1$ dimensional world volume.  This theory exhibits
interesting properties which did not arise for the D3-D5
intersection.  Unlike the D3-D5 intersection, open strings on both
stacks of branes remain coupled as $\alpha^{\prime} \rightarrow
0$.  The resulting theory is described by a linear sigma model on
two intersecting world volumes.  The classical Higgs branch of
this theory has an interpretation as a smooth resolution of the
intersection to the holomorphic curve $wy\sim c\alpha^{\prime}$,
where $w= X^2 + i X^3$ and $y = X^4 + iX^5$. However, due to the
two-dimensional nature of the fields which parameterize these
curves the quantum vacuum spreads out over the entire
classical Higgs branch.

As a result of the spreading over the Higgs branch,  it has been
argued that a fully localized supergravity solution for this
D3-brane intersection does not exist
\cite{marolfpeet,Gomberoff,peet}. Obtaining a closed string
description of this defect CFT would therefore seem to be
difficult. From the point of view of the linear sigma model
description, a  holographic equivalence with a closed string
background would seem to require a target space with a singular
boundary. Nevertheless, there is a limit in which a holographic
duality be found relating fluctuations in an $AdS$ background to
operators in the linear sigma model. One simply takes the number
of D3-branes, $N$, in the first stack to infinity, keeping $g_sN$
and the number of D3-branes in the second stack, $N^{\prime}$,
fixed. In this limit, the 't~Hooft coupling of the gauge theory on
the second stack, $\lambda^{\prime} = g_s N^{\prime}$, vanishes.
Thus the open strings with all endpoints on the second stack
decouple, and one is left with a four-dimensional CFT with a
codimension two defect.  The defect breaks half of the original
\mbox{$\N=4$}, $d=4$ supersymmetry, leaving eight real
supercharges realizing a two-dimensional $(4,4)$ supersymmetry
algebra. The conformal symmetry of the theory is a global $SL(2,R)
\times SL(2,R)$,  corresponding to a subgroup of the
four-dimensional conformal symmetries.  The degrees of freedom at
the impurity are a $(4,4)$ hypermultiplet arising from the open
strings connecting the orthogonal stacks of D3-branes.

In the limit described above, the holographic dual is obtained by
focusing on the near horizon region for the first stack of
D3-branes, while treating the second stack as a probe.  The result
is an $AdS_5 \times S^5$ background with $N^{\prime}$ probe
D3-branes wrapping an $AdS_3\times S^1$ subspace. This embedding
was shown to be supersymmetric in \cite{SkenderisTaylor}.
We will demonstrate below that there is a one complex parameter family of
such
embeddings, corresponding to the holomorphic curves $wy\sim c$,
all of which preserve a set of isometries corresponding to the
super-conformal group. In the spirit
of \cite{KR}, holographic duality is conjectured to act ``twice''.
First there is the standard AdS/CFT duality relating closed
strings in $AdS_5 \times S^5$ to operators in ${\cal N}=4$ super
Yang-Mills theory. Second, there is a duality relating open
strings on the probe D$3^{\prime}$ wrapping $AdS_3\times S^1$ to
operators localized on the $1+1$ dimensional defect.

One of the original motivations to search for holographic
dualities for defect conformal field theories \cite{KR} is that
such a duality might imply the localization of gravity on branes
in string theory.  In the context of a brane wrapping an $AdS_3$
geometry embedded inside $AdS_5$, localization of gravity would
indicate the existence of a Virasoro algebra in the dual CFT,
through a Brown-Henneaux mechanism \cite{BrownHenneaux}. We do not
find any evidence for the existence of a Virasoro algebra in the
conformal field theory. Although this theory has a $(4,4)$
superconformal algebra, only the finite part of the algebra is
realized in any obvious way. Roughly speaking, the $(4,4)$
superconformal algebra is the common intersection of two ${\cal N}
= 4, d=4$ superconformal algebras, both of which are finite. The
even part of the superconformal group is $SL(2,R) \times SL(2,R)
\times SU(2)_L \times SU(2)_R \times U(1)$,  which is also
realized as an isometry of the $AdS_5$ background which preserves
the probe embedding. Enhancement to the usual infinite dimensional
algebra would require the existence of a decoupled two-dimensional
sector. Correctly addressing this issue would require going beyond
the probe limit and studying the back reaction of the
D$3^{\prime}$-branes on the $AdS_5\times S^5$ geometry as well as
gaining a deeper understanding of the dynamics of the defect CFT.

The action for the D3-D3 intersection is most easily and elegantly
constructed in $(2,2)$ superspace.  Although it may seem unusual
to write the ${\cal N} =4, d=4$ components of the action in
$(2,2)$ superspace, this is actually quite natural because the
four-dimensional supersymmetries are broken by couplings to the
defect hypermultiplet.  In writing this action,  we will not take
the limit which decouples one stack of D3-branes. With the help of
the manifest chirality of $(2,2)$ superspace we are able to find
an argument for the absence of quantum corrections to the combined
2d/4d actions, which implies that the theory remains conformal
upon quantization.  Although this theory has two-dimensional
fields coupled to gauge fields, the gauge couplings couplings are
exactly marginal due to the four-dimensional nature of the gauge
fields.

We give a detailed dictionary between Kaluza-Klein fluctuations on
the probe D3-brane and operators localized on the defect. Of
particular interest will be a certain subset of the fluctuations
which describe the embedding of the probe inside  $AdS_5$. This
subset is dual to operators containing defect scalar fields, which
appear without any derivative or vertex operator structure. Due to
strong infrared effects in two dimensions, these fields are not
conformal fields associated to states in the Hilbert space. From
the point of view of the probe-supergravity system, there is at
first sight nothing unusual about these fluctuations.  However
upon applying the usual $AdS_3$/CFT$_2$ rules to compute the dual
two-point correlator,  one finds identically zero due to extra
surface terms in the probe action.  Thus there is no clear
interpretation of these fluctuations as sources for the generating
function of the CFT. We shall find however that the bottom of the
Kaluza-Klein tower for these fluctuations (with appropriate
boundary conditions) parameterizes the aforementioned holomorphic
embedding of the probe inside $AdS_5$. While the interpretation of
this fluctuation as a source is unclear, it nevertheless labels
points on the classical Higgs branch. Since the infrared dynamics
of two dimensions implies that the vacuum is spread out over the
entire Higgs branch, one should in principle sum over holomorphic
embeddings when performing computations in the $AdS$ background.

The fluctuations of the probe $S^1$ embedding inside $S^5$ satisfy
the Breitenlohner-Freedman bound despite the lack of topological
stability. These fluctuations are dual to a multiplet of scalar
operators with defect fermion pairs which we identify with BPS
superconformal primaries localized at the intersection. We also
find fluctuations of the probe embedding inside $AdS_5$ which are
dual to descendants of these operators. Remarkably, the AdS
computation of the corresponding correlators, which is valid for
large 't Hooft coupling $\lambda$, shows no dependence on
$\lambda$. We also study perturbative quantum corrections to the
two-point function of the BPS primary operators and find that such
corrections are absent at order $g_{YM}^2$. Together with the
$AdS$ strong coupling result, this suggests the existence of a
non-renormalization theorem.

The paper is organized as follows. In section 2 we present the
D3-brane setup, its near horizon isometries and the superconformal
algebra. In section 3 we obtain the spectrum of low-energy
fluctuations about the probe geometry. In section 4 we show that
the n-point functions associated with these fluctuations are
independent of the 't~Hooft coupling,  at least when the 't Hooft
coupling is large.  Moreover we show that the classical action for
probe fluctuations dual operators parameterizing the classical
Higgs branch does correspond to a power law two-point function. In
section 5 we study the field theory associated with the D3-brane
intersection. We obtain the action using $(2,2)$ superspace for
both the defect and four-dimensional components.  In section 6 we
derive the fluctuation-operator dictionary for the conjectured
AdS/CFT correspondence. In section 7 we demonstrate that two-point
functions of the BPS primary operators identified in section 6 do
not receive any radiative corrections to order $g^2$, thus
providing evidence for a non-renormalization theorem. We conclude
in section 8 by presenting some open questions. There is a series
of appendices containing further details of the calculations. In
particular in appendix E we give an argument for quantum conformal
invariance of the defect CFT which holds to all orders in
perturbation theory.

\section{Holography for intersecting D3--branes}
\setcounter{equation}{0}

\subsection{The configuration}

We are interested in the conformal field theory describing the low
energy limit of a stack of $N$ D3-branes in the $x^0,x^1,x^2,x^3$
directions intersecting  another stack of $N^{\prime}$ D3-branes
in the $x^0,x^1,x^4,x^5$ directions, as indicated in the following
table:
\bigskip
\begin{table}[!h]
\begin{center}
\begin{tabular}{|l|c|c|c|c|c|c|c|c|c|c|}
\hline
& 0 & 1 & 2 & 3 & 4 & 5 & 6 & 7 & 8 & 9 \\
\hline
D3    & X & X & X & X &   &   &   &   &   &   \\
\hline
D$3'$ & X & X &   &   & X & X &   &   &   &   \\
\hline
\end{tabular}
\end{center}
\end{table}

This intersection preserves $8$ supersymmetries.  The massless
open string degrees of freedom correspond to a pair of ${\cal
N}=4$ super-Yang-Mills multiplets coupled to a bifundamental
$(4,4)$ hypermultiplet localized at the $1+1$ dimensional
intersection. The coupling is such that a two-dimensional $(4,4)$
supersymmetry is preserved. We shall study the holographic
description of this system in a limit in which one of the ${\cal
N}=4$ multiplets decouples, leaving a single ${\cal N}=4$
multiplet coupled to a $(4,4)$ hypermultiplet at a $1+1$
dimensional defect. This decoupling is achieved by scaling
$N\rightarrow\infty$ while keeping $g_s N\sim g^2_{YM} N$ and
$N^{\prime}$ fixed.  This is the usual 't~Hooft limit for the
gauge theory describing the $N$ D3-branes. For $\lambda \equiv
g_{YM}^2N \gg 1$ one may replace the $N$ D3-branes by the geometry
$AdS_5\times S^5$, according to the usual AdS/CFT correspondance.
On the other hand, the 't~Hooft coupling for the $N^{\prime}$
orthogonal D3-branes is $\lambda'= g_sN' = \lambda N'/N$ which
vanishes in the above limit. For large $\lambda$, one may treat
these branes as a probe of an $AdS_5\times S^5$ geometry.

\begin{figure}%[!h]
\begin{center}
\includegraphics[scale=0.9]{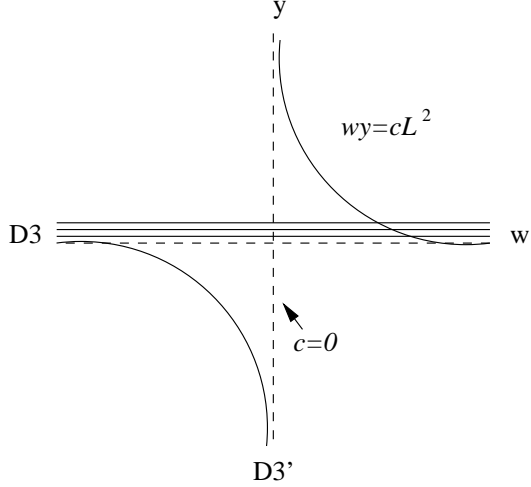}
\caption{Holomorphic curve $wy=cL^2$.} \label{fig1}
\end{center}
\end{figure}

We now demonstrate the existence of a one complex parameter family
of $AdS_3 \times S^1$ embeddings of the probe D$3^{\prime}$-branes
in the $AdS_5\times S^5$ background. Consider first the geometry
of the stack of $N$ D3-branes before taking the near horizon
limit. The D3 metric is given by
\begin{align} ds^2 = \left(1+\frac{L^4}{r^4}\right)^{-\frac{1}{2}}
(-dt^2 + dx_1^2+dx_2^2+dx_3^2)+
\left(1+\frac{L^4}{r^4}\right)^{\frac{1}{2}} (dx_4^2+\cdots
+dx_9^2)
\end{align}
We will choose a static gauge in which the world volume coordinates of the
probe are identified with $t,x^1,x^4,x^5$. Defining $w=x^2+ix^3$ and
$y=x^4+ix^5$ the probe is taken to lie on the surface defined by $wy=cL^2$
and
$x^6=x^7=x^8=x^9=0$. Here $c$ is an arbitrary complex number.
When $c=0$ we have $w=0$ and the probe sits at the origin of the space
transverse to it's world volume. For $c\neq 0$ the probe sits on a
holomorphic
curve embedded into the space spanned by $x^{2,3,4,5}$ (see figure
\ref{fig1}).  With this choice of embedding the induced metric on the probe
world volume is, \beq ds^2_{probe}= h^{-1/2}\left(-dt^2+dx_1^2\right)+
h^{1/2}\left(1+\frac{|c|^2L^4}{(|y|^2)^2}h^{-1}\right)dyd\bar{y}
\label{probemet}
\eeq
where $h=1+L^4/(|y|^2)^2$ is the harmonic function appearing in the
background
geometry evaluated at the position of the probe.

In the near horizon limit, $L/r\gg 1$, the D3-brane geometry becomes $AdS_5
\times S^5$,
\begin{align}
ds^2_{AdS_5 \times S^5} =
&\frac{L^2}{u^2}(-dt^2+dx_1^2+dx_2^2+dx_3^2+du^2) \nonumber\\ &+
L^2\left(d\phi^2+ s_\phi^2d\theta^2 + s_\phi^2 s_\theta^2 d\rho^2
+ s_\phi^2 s_\theta^2 s_\rho^2 d\varphi^2 + s_\phi^2 s_\theta^2
s_\rho^2 s_\varphi^2 d\xi^2\right) \end{align} where $u\equiv
\frac{L^2}{r}$ and we have defined angular variables $\phi,
\theta, \rho, \varphi, \xi$ via
\beqa
x^4 &=& r s_\phi s_\theta s_\rho s_\varphi s_\xi \,\,\,\,\,\,\,\,\,
x^5 =r s_\phi s_\theta s_\rho s_\varphi c_\xi \nonumber \\
x^6 &=& r s_\phi s_\theta s_\rho c_\varphi \,\,\,\,\,\,\,\,\,\,\,\,\,\,
x^7 = r s_\phi s_\theta c_\rho \nonumber \\
x^8 &=& r s_\phi c_\theta
\,\,\,\,\,\,\,\,\,\,\,\,\,\,\,\,\,\,\,\,\,\,\,\,\, x^9 = r c_\phi
\label{sphere} \eeqa where $s_\phi = \sin\phi$, $c_\phi =
\cos\phi$ etc. It is instructive to consider this limit from the
point of view of the probe metric. One can easily show that in the
near horizon region the induced metric on the probe becomes, \beq
ds^2_{probe}
=\frac{\tilde{L}^2}{\tilde{u}^2}\left(-dt^2+dx_1^2+d\tilde{u}^2\right)
+\tilde{L}d\xi^2 \label{nearprobe} \eeq where
$\tilde{L}^2=L^2(1+|c|^2)$ and \beq \tilde{u} =
\frac{\tilde{L}}{L}u|_{x^6,x^7,x^8,x^9=0} \,. \label{hatu} \eeq
One immediately recognizes eqn.~\reef{nearprobe} as the metric on
$AdS_3\times S^1$ with radius of curvature~$\tilde{L}$. The probe
is sitting at $\phi=\theta=\rho=\varphi=\pi/2$ and thus wraps a
circle of maximal radius inside the $S^5$. For the special case
$c=0$ the curvature is the same as that of the ambient
$AdS_5\times S^5$ geometry. For $c\neq 0$ however the effective
cosmological constant on the probe differs from that of the bulk
of $AdS_5$. This is reminiscent of the D3-D5 system studied in
ref.~\cite{KR} in which D5-brane probes were wrapped on an
$AdS_4\times S^2$ slice of the full geometry. In that case probe
D5-branes were able to have effective cosmological constants which
differed from the bulk when some of the D5-branes ended on the
D3-branes~\cite{KR}. Here the probe D$3^{\prime}$-branes cannot
end on the D3-branes however one of the probe branes can `merge'
with one of the $N$ D3-branes and form a holomorphic  curve. It is
this holomorphic curve that is parameterized by $c$. Notice that
$c$ also parameterizes a family of $AdS_3$ spaces and therefore we
expect that this deformation preserves the conformal invariance of
the dual field theory. It is interesting that while supersymmetry
allows for {\it any} holomorphic curve of the form $wy^l =
cL^{l+1}$~\cite{calibrate} only for $l=1$ is the $AdS_3$ geometry
and hence conformal invariance preserved. For the majority of this
paper we will restrict our attention to the case $c=0$ however we
will return to the general case when we discuss the classical
Higgs branch of the dual defect conformal field theory.

The boundary of the embedded $AdS_3$ is an
$\mathbb{R}^2$ at $\tilde{u}=0$, and lies within the $\mathbb{R}^4$
boundary of $AdS_5$. This embedding is indeed supersymmetric, as
was verified for $c=0$  in \cite{SkenderisTaylor}. Thus this configuration
is
stable despite the fact that the $S^1$ is contractible within the
$S^5$.  As we will see shortly,  the naively unstable modes
associated with contracting the $S^1$ satisfy the
Breitenlohner-Freedman bound \cite{BF} for scalars in $AdS_3$, and
therefore do not lead to an instability.

\begin{figure}[!ht]
\begin{center}
\includegraphics{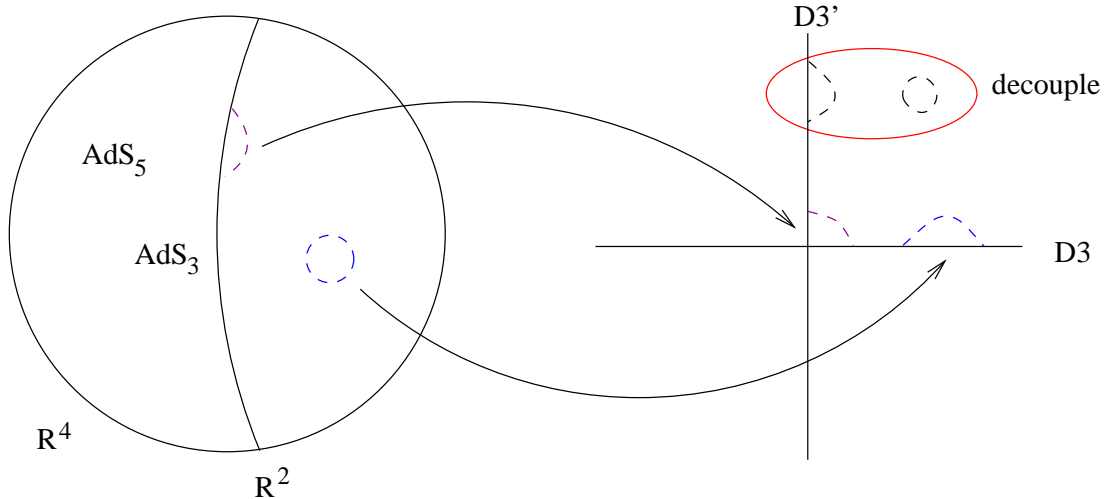}
\caption{AdS/CFT duality for an Impurity CFT. The duality acts
twice.  Once for the IIB supergravity on $AdS_5 \times S^5$, and
once for DBI theory on $AdS_3 \times S^1$.}  \label{fig2}
\end{center}
\end{figure}

Following the arguments of \cite{KR,DFO} we propose that AdS/CFT
duality ``acts twice'' in the background with an $AdS_3$ brane
embedded in $AdS_5$. This means that the closed strings on $AdS_5$
should be dual to ${\cal N} =4$ $SU(N)$ super Yang-Mills theory on
$\mathbb{R}^4$,  while open string modes on the probe $AdS_3$
brane should be dual to the fundamental $(4,4)$ hypermultiplet on
the $\mathbb{R}^2$ defect (see figure \ref{fig2}). Interactions between the
defect hypermultiplet and the bulk $\N=4$ fields should correspond
to couplings between open strings on the probe D3-brane and closed
strings in  $AdS_5 \times S^5$. For large 't~Hooft coupling,
the generating function for correlation functions of the defect
CFT should be given by the classical action of IIB supergravity on
$AdS_5 \times S^5$ coupled to a Dirac-Born-Infeld theory on
$AdS_3\times S^1$,  with suitable conditions on the behaviour of
fields at the boundary of $AdS_5$ and $AdS_3$.

\subsection{Isometries}

In the absence of the probe D3-branes,  the isometry group of the
$AdS_5 \times S^5$ background is $SO(2,4) \times SO(6)$.  The
$SO(2,4)$ component acts as conformal transformations on the
boundary of $AdS_5$,  while the $SO(6) \sim SU(4)$ isometry of
$S^5$ is the R-symmetry of four-dimensional ${\N =4}$ super
Yang-Mills theory,  under which the six real scalars
$X^{4,5,6,7,8,9}$ transform in the vector representation.

In the presence of the probe D3-brane, the $AdS_5 \times S^5$
isometries are broken to the subgroup which leaves the embedding
equations of the probe invariant:
\begin{align}
\!\!\!SO(2,4)  \times SU(4)  \rightarrow SL(2,R)\times SL(2,R)
\times U(1) \times SU(2)_L \times SU(2)_R \times U(1)
\label{unb}\end{align} Out of the $SO(2,4)$ isometry of $AdS_5$
only $SO(2,2) \times U(1) \simeq SL(2,R) \times SL(2,R) \times U(1)$ is
preserved. The $SO(2,2)\simeq SL(2,R) \times SL(2,R)$ factor is the isometry
group of $AdS_3$, while the $U(1)$ factor acts as a phase rotation of the
complex coordinates $w = X^2 + iX^3$. Out of the $SO(6) \simeq SU(4)$ 
isometry
of $S^5$, only $SO(4) \times U(1) \simeq SU(2)_L \times SU(2)_R \times U(1)$
is preserved.  The $U(1)$ factor here acts as phase rotation of the complex
coordinate $y = X^4+iX^5$, which rotates the $S^1$ of the probe worldvolume.
The $SO(4)$ component acts on the coordinates $X^{6,7,8,9}$. As we shall see
in section \ref{superalg}, only a certain combination of the two $U(1)$
factors in (\ref{unb}) enters the superconformal algebra. The even part of 
the
superconformal group is $SL(2,R) \times SL(2,R) \times SU(2)_L \times 
SU(2)_R
\times U(1)$.

\subsection{The superconformal algebra} \label{superalg}

The D3-D3 intersection has a $(4,4)$ superconformal group whose
even part is $SL(2,R) \times SL(2,R) \times SU(2)_L \times SU(2)_R
\times U(1)$.  We emphasize that this system does not give a
standard $(4,4)$ superconformal algebra.  Because of the couplings
between two and four-dimensional fields, the algebra does not
factorize into left and right moving parts. Neither an infinite
Virasoro algebra nor an affine Kac-Moody algebra are realized in
any obvious way. The superconformal algebra for the D3-D3 system
should be thought of as a common ``intersection'' of two ${\cal
N}=4, d=4$ superconformal algebras, both of which are finite. If
there is a hidden affine algebra, it should arise via some
dynamics which gives a decoupled two-dimensional sector,  for
which we presently have no evidence.

For comparative purposes,  it is helpful to first review the
situation for more familiar two-dimensional $(4,4)$ theories with
vector multiplets and hypermultiplets,  such as those considered
in \cite{wittenhiggs}. These theories may have classical Higgs and
Coulomb branches which meet at a singularity of the moduli space.
For finite coupling, quantum states spread out over both the Higgs
and Coulomb branches. However in the infrared (or strong coupling)
limit, one obtains a separate $(4,4)$ CFT on the Higgs and Coulomb
branches \cite{wittenhiggs}. One argument for the decoupling of
the Higgs and Coulomb branches is that the $(4,4)$ superconformal
algebra contains an $SU(2)_l \times SU(2)_r$ R-symmetry with a
different origin in the original $SU(2)_L \times SU(2)_R \times
SU(2)$ R-symmetry depending on whether one is on the Higgs branch
or Coulomb branch. The CFT scalars must be uncharged under the
R-symmetries. This means for example that the original $SU(2)_L
\times SU(2)_R$ factor may be the R-symmetry of the CFT on the
Higgs branch but not the Coulomb branch.

For the linear sigma model describing the D3-D3 intersection, a
$(4,4)$ super-conformal theory arises only on the Higgs branch,
which is parameterized by the two-dimensional scalars of the
defect hypermultiplet.  On the Coulomb branch,  the orthogonal
D3-branes are separated by amounts characterized by the VEV's of
four-dimensional scalar fields.  One obtains a CFT on the Higgs
branch without flowing to the IR, since the gauge fields propagate
in four dimensions and the gauge coupling is exactly
marginal.\footnote{We give field theoretic arguments for quantum
conformal invariance in Appendix E.} Furthermore scalar degrees
of freedom of the CFT may carry R-charges, since the R-currents do
not break up into purely left and right moving parts.  Of course
the scalars of the defect hypermultiplet must still be uncharged
under the R-symmetries, since for $g_{YM}=0$ the free $(4,4)$
hypermultiplet realizes a conventional two-dimensional $(4,4)$
CFT.  However the four-dimensional scalar fields, which are not
decoupled at finite $g_{YM}$, transform non-trivially under the
R-symmetries of the defect CFT.

In more familiar considerations of the $AdS_3/CFT_2$ duality, the
full Virasoro algebra is realized in terms of diffeomorphisms that
leave the form of the metric invariant asymptotically, near the
boundary of $AdS_3$ \cite{BrownHenneaux}.  Of these
diffeomorphisms, the finite $SL(2,R) \times SL(2,R)$ subalgebra is
realized as an exact isometry.  However the three-dimensional
diffeomorphisms which are asymptotic isometries of $AdS_3$, and
correspond to higher-order Virasoro generators, do not have an
extension into the bulk which leave the $AdS_5$ metric
asymptotically invariant.  The existence of a Virasoro algebra
seems to require localized gravity on $AdS_3$.  This could only be
seen through a consideration of the back-reaction.  In the defect
CFT,  the two-dimensional conformal algebra contains only those
generators which can be extended to conformal transformations of
the four-dimensional parts of the world volume, namely $L_{-1},
L_0, L_1,\tilde L_{-1},\tilde L_0$ and $\tilde L_1$.

The `global' $(4,4)$ superconformal algebra of defect CFT  gives
relations between the dimensions and R-charges of BPS operators.
We will later find that these relations are consistent with the
spectrum of fluctuations in the probe-$AdS$ background. To
construct the relevant part of the algebra, it is helpful to note
that the algebra should be a subgroup of an ${\cal N} =4, d=4$
superconformal algebra (or actually an unbroken intersection of
two such algebras).

Let us start by writing down the relevant part of the ${\cal N}
=4, d=4$ superconformal algebra for the D3-branes in the $0123$
directions. The supersymmetry generators are $Q_{\alpha}^a$, where
$\alpha =1,2$ is a spinor index and $a=1, \cdots,4$ is an index in
the representation ${\bf 4}$ of the $SU(4)$ R-symmetry.  The
special superconformal generators are $S_{\beta b}$ which are in
the ${\bf 4}^*$ representation of $SU(4)$. The relevant part of
the ${\cal N} =4, d=4$ algebra is then
\begin{align}
\{Q^a_{\alpha},S_{\beta b} \} =
\epsilon_{\alpha\beta}({\delta_b}^a D + 4 J^A (T_A)^a_b) +
\frac{1}{2}{\delta_b}^a L_{\mu\nu}\sigma^{\mu\nu}_{\alpha\beta}\,
, \label{alg}
\end{align}
where $D$ is the dilation operator,  $J^A$ are the operators
generating $SU(4)$, and $L_{\mu\nu}$ are the generators of
four-dimensional Lorentz transformations.  The matrices $(T_
A)^a_b$ generate the fundamental representation of $SU(4)$, and
are normalized such that $\Tr(T^AT^B) = \frac{1}{2} \delta^{AB}$.

A $(4,4)$ supersymmetry sub-algebra is generated by the
supercharges $Q^a_1 \equiv Q^a_+$ with $a=1,2$, and $Q^a_2\equiv
Q^a_-$ with $a = 3,4$, on which an $SU(2)_L \times SU(2)_R \times
U(1)$ subgroup of the orginal $SU(4)$ R-symmetry acts.  The
embedding of the $SU(2)_L \times SU(2)_R \times U(1)$ generators
in $SU(4)$ is as follows:
\begin{align}SU(2)_L: \begin{pmatrix} \frac{1}{2}\sigma^A & 0\cr 0 & 0
\end{pmatrix}, \quad
SU(2)_R: \begin{pmatrix} 0 & 0 \cr 0 & \frac{1}{2}\sigma ^B
\end{pmatrix},\quad U(1): \frac{1}{\sqrt{8}} \begin{pmatrix}-I & 0 \cr 0 & I
\end{pmatrix} \,.\end{align}
The unbroken $SU(2)_L \times SU(2)_R$ R-symmetry corresponds to
rotations in the directions $6,7,8,9$ transverse to both stacks of
D3-branes, while the unbroken $U(1)$ describes rotation in the
$45$ plane. These symmetries act on adjoint scalars.  Since the
R-currents of the CFT do not break up into left and right moving
parts, there is no requirement that four-dimensional scalars are
uncharged under R-symmetries.  We shall call the generator of
rotations in the $45$ plane $J_{45}$, and normalize it such that
the supercharges $Q^a_\pm$ have $J_{45}$ eigenvalue $\pm 1/2$. The
special superconformal generators of the $(4,4)$ sub-algebra are
$S_{b2} \equiv S_{b-}$ with $b=1,2$ and $S_{b1} \equiv S_{b+}$
with $b=3,4$.    The term in the $(4,4)$ algebra inherited from
(\ref{alg}) is then
\begin{align}
\{Q^a_+,S_{b-} \} &= {\delta_b}^a D + 2J^L_A(\sigma^A)^a_b +
\delta^a_b J_{45} + \delta^a_b L_{01} + \delta^a_b L_{23}\, , \label{one} \\
\{Q^a_-,S_{b+} \} &=  -\delta_b^a D - 2J^R_A(\sigma^A)^a_b +
\delta^a_b J_{45} + \delta^a_b L_{01} + \delta^a_b L_{23} \,.
\label{two}
\end{align}
The unbroken Lorentz generators are $L_{01}$ and $L_{23}$. Note
that from a two-dimensional point of view,  the Lorentz
transformations are generated by $L_{01}$, whereas $L_{23}$ is an
R-symmetry.

\begin{figure}[!ht]
\begin{center}
\includegraphics{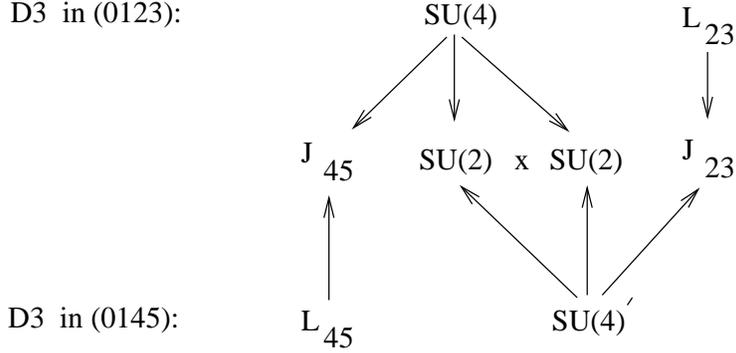}
\caption{Decomposition of the two $SU(4)$ R-symmetries.}  \label{decompo}
\end{center}
\end{figure}

For the orthogonal D3-branes spanning $0,1,4,5$, rotations in the
$45$ plane are Lorentz generators $L_{45}$ rather than a subgroup
of $SU(4)$. The rotations in the $23$ plane are an unbroken $U(1)$
part of the $SU(4)$ R-symmetry rather than a Lorentz
transformation.  This distinction is illustrated in figure
\ref{decompo}.

\noindent From the two-dimensional point of view, both $23$ and
$45$ rotations are $U(1)$ R-symmetries. If we write $L_{23} =
J_{23}, \, L_{45} = J_{45}$ and define ${\cal J} = J_{23}+J_{45}$,
then the terms (\ref{one}) and (\ref{two}) become
\begin{align}
\{Q^a_+,S_{b-} \} &= {\delta_b}^a (L_{01}+D) + 2J^L_A(\sigma^A)^a_b
+
\delta^a_b {\cal J}\, , \label{thealg1} \\
\{Q^a_-,S_{b+} \} &=  \delta_b^a (L_{01}-D)  -
2J^R_A(\sigma^A)^a_b + \delta^a_b {\cal J}
\,.\label{thealg}\end{align} which are applicable to {\it both}
stacks of D3-branes.  This forms part of the $(4,4)$
superconformal algebra of the full D3-D3 system. The charge ${\cal
J}$ plays a somewhat unusual role. From the point of view of the
bulk four-dimensional fields, ${\cal J}$ is a combination of an
R-symmetry and a Lorentz symmetry, under which the preserved
supercharges are invariant. As we will see later, the fields
localized at the two-dimensional intersection are not charged
under ${\cal J}$. Upon decoupling the four-dimensional fields by
taking $g = 0$, the two-dimensional sector becomes a free $(4,4)$
superconformal theory with an affine $SU(2)_L \times SU(2)_R$
R-symmetry. However, for $g_{YM} \ne 0$, the algebra does not
factorize into left and right moving parts.

The algebra (\ref{thealg1},~\ref{thealg}) determines the dimensions of
the BPS superconformal primary operators, which are annihilated by
all the $S$'s and some of the $Q$'s.  The bounds on dimensions due
to the superconformal algebra are best obtained in Euclidean
space.  The Euclidean $(4,4)$ algebra of the defect CFT  contains
the terms
\begin{align}
\{ {\cal Q}^a_{1/2}, {\cal Q}_{1/2}^{b\dagger} \} = 2\delta^a_b
L_0 + 2J^L_A ({\sigma^A})^a_b + \delta^a_b {\cal
J}\, , \label{left} \\
\{ \tilde {\cal Q}^a_{1/2}, \tilde {\cal Q}_{1/2}^{b\dagger} \} =
2\delta^a_b \tilde L_0 + 2J^R_A  ({\sigma^A})^a_b  - \delta^a_b
{\cal J}\, . \label{right}
\end{align}
For $a=b$, the left hand side of (\ref{left}) and (\ref{right})
are positive operators, leading to the bounds
\begin{align}
h + j^L_3 + \textstyle\frac{1}{2}{\cal J} \geq 0 \,,\label{first}\\
h - j^L_3 + \textstyle\frac{1}{2}{\cal J} \geq 0 \,,\label{second}\\
\tilde h + j^R_3 - \textstyle\frac{1}{2}{\cal J} \geq 0\,, \label{third}\\
\tilde h - j^R_3 - \textstyle\frac{1}{2}{\cal J} \geq 0\,, \label{fourth}
\end{align}
some of which are saturated by the BPS super-conformal primaries.
As always, the dimensions are $\Delta = h + \tilde h$,  with
$h=\tilde h$ for scalar operators.

\section{Fluctuations in the probe-supergravity background}
\setcounter{equation}{0}

Following the conjecture put forth in \cite{KR}  and elaborated
upon in \cite{DFO},  we expect the holographic duals of defect
operators localized on the intersection are open strings on the
probe D$3^{\prime}$, whose world volume is an $AdS_3 \times S^1$
submanifold of $AdS_5 \times S^5$.  The operators with protected
conformal dimensions should be dual to probe Kaluza-Klein
excitations at ``sub-stringy'' energies, $m^2 \ll \lambda/L^2$. In
this section we shall find the mass spectra of these excitations.
Later we will find this spectrum to be consistent with the
dimensions of operators localized on the intersection.

\subsection{The probe-supergravity system}

The full action describing physics of the background as well as
the probe is given by \beq S_{bulk}=S_{IIB}+S_{DBI}+S_{WZ} \,.
\labell{totact} \eeq The contribution of the bulk supergravity
piece of the action in Einstein frame is \beq
S_{IIB}=\frac{1}{2\kappa^2}\int
d^{10}x\sqrt{-g}\left(R-\frac{1}{2}e^{2\Phi} (\partial\Phi)^2
+\cdots\right) \labell{IIB} \eeq where
$2\kappa^2=(2\pi)^7g_s^2l_s^8$. The dynamics of the probe D3-brane
is given by a Dirac-Born-Infeld term and a Wess-Zumino term
\cite{leigh}, \beq S_{DBI} + S_{WZ}= -T_{D3}\int d^4\sigma
e^{-\Phi}\sqrt{-\det\left(g_{ab}^{PB} + e^{-\Phi/2}{\cal
F}_{ab}\right) } \,+\,T_{D3}\int C^{(4)}_{PB} \,. \labell{d3act}
\eeq The metric $g_{ab}^{PB}$ is the pull back of the bulk $AdS_5
\times S^5$ metric to the world volume of the probe,  while
$C^{(4)}_{PB}$ is the pull back of the bulk Ramond-Ramond four
form.

We work in a static gauge where the world volume coordinates of
the brane are identified with the space time coordinates by
$\sigma^{a} \sim x^0,x^1,u,\xi $. With this identification the DBI
action is \beq S_{DBI}=-T_{D3} \int
d^4\sigma\sqrt{-\det\left(g_{ab}+
\partial_aZ^i\partial_bZ^jg_{ij}+e^{-\Phi/2}{\cal F}_{ab} +
2g_{ai}\partial_bZ^i\right)} \labell{actt} \eeq where $i,j$ label
the transverse directions to the probe and the scalars $Z^i$
represent the fluctuations of the transverse scalars $X^2, X^3,
\theta, \phi, \rho, \varphi$. Also, ${\cal F}_{ab} = B_{ab}+ 2\pi
l_s^2F_{ab}$ is the total world volume field strength. Henceforth we will
only consider the open string fluctuations on the probe and thus drop terms
involving closed string fields\footnote{Such terms encode the physics of
operators
in the bulk of the dual ${\cal N} =4$ theory restricted to the
defect.}$B_{ab}$
and $g_{ai}$. To quadratic order in fluctuations, the action takes the form
\begin{align}
S_{DBI}&=-T_{D3}L^4 \int d^4\sigma \sqrt{\bar g_4}
(1-\frac{1}{2}\phi^2 - \frac{1}{2} \theta^2 -\frac{1}{2} \rho^2
-\frac{1}{2}\varphi^2 +
\frac{1}{2}\partial_a\theta\partial^a\theta +
\frac{1}{2}\partial_a\phi\partial^a\phi \nonumber\\
&+ \frac{1}{2}\partial_a\rho\partial^a \rho +
\frac{1}{2}\partial_a\varphi\partial^a\varphi +
\frac{1}{2u^2}\partial_a X^2\partial^aX^2
+\frac{1}{2u^2}\partial_aX^3\partial^aX^3 +\frac{1}{4}(2\pi l_s^2)^2
F_{ab} F^{ab} ) \label{quadBI}
\end{align}
where $\bar{g}_4$ is the determinant of the rescaled $AdS_3\times S^1$
metric $\bar g^4_{ab}$ given by
\begin{align}
d\bar s^2 = \frac{1}{u^2}(- dt^2 + dx^2_1 +du^2 ) + d\xi^2.
\end{align}

To obtain the Wess-Zumino term $S_{WZ}$ we require the pull back
of the bulk RR-four form to the probe: \beqa C^{PB}_{abcd}&=&
C_{abcd} + 4\prt_{[a}Z^iC_{bcd]i}
+6\prt_{[a}Z^i\prt_{b}Z^{j}C_{cd]ij} \nonumber\\
&+&4 \prt_{[a}Z^i\prt_{b}Z^{j}\prt_{c}Z^kC_{d]ijk} +
\prt_{[a}Z^i\prt_{b}Z^{j}\prt_{c}Z^k\prt_{d]}Z^lC_{ijkl} \,. \eeqa
In the $AdS_5 \times S^5$ background,  one can choose a gauge in
which
\begin{align}
C^{(4)}_{0123} = \frac{L^4}{u^4} \label{comp}
\end{align}
while the remaining components,  which are determined by the self
duality of $dC^{(4)}$, contribute only to terms in the pull back
with more than two $\partial Z$'s.  We do not need such terms to
obtain the fluctuation spectrum.  The quadratic term arising from
(\ref{comp}) is \beq
C_{PB}^{(4)}=\left(\prt_{u}X^2\prt_{\xi}X^3-\prt_{u}X^3\prt_{\xi}X^2\right)
C_{0123}dt \wedge dx^1\wedge du\wedge d\xi \, . \eeq The Wess-Zumino
action is then
\begin{align}
S_{WZ} = T_{D3}L^4 \int d^4\sigma \frac{1}{u^4}(\partial_u X^2
\partial_{\xi}X^3 - \partial_u X^3 \partial_\xi X^2) \,.
\label{quadWZ}
\end{align}

\subsection{$S^1$ fluctuations inside $S^5$}

{}From eqn.~\reef{quadBI} one can see that the angular
fluctuations $\theta, \phi, \rho$ and $\varphi$ are minimally
coupled scalars on $AdS_3\times S^1$. Interestingly they have $m^2
=-1$ which, although negative, satisfies (saturates) the
Breitenlohner-Freedman bound $m^2 \ge -d^2/4$, where $d=2$ for
$AdS_3$. Expanding in Fourier modes on $S^1$, {\it i.e.,} $\theta
= \theta_l e^{il\xi}$ the Kaluza-Klein modes of these scalars have
$m^2 = -1 + l^2$. This leads to a spectrum of conformal dimensions
of dual defect operators given by $\Delta_\pm = \frac{d}{2} \pm
\sqrt{\frac{d^2}{4} + m^2} = 1 \pm l$, where $d=2$. For $l>0$ one
should choose the positive branch for unitarity, while for $l<0$
one should choose the negative branch. To leading order in
fluctuations of the $S^1$ embedding we see that
\begin{align}
x^6 &= -r\varphi \nonumber \\ x^7 &= -r\rho \nonumber \\ x^8 &=
-r\theta \nonumber \\ x^9 &= -r\phi \,. \end{align} Thus the
angular variables $\phi,\theta,\rho$ and $\varphi$ belong to a
$(\frac{1}{2},\frac{1}{2})$ multiplet of $SU(2)_L \times SU(2)_R$.
Moreover, these fluctuations have $J_{23}=0$ and $J_{45}=l$ such
that the $U(1)$ charge appearing in the algebra (\ref{left},~\ref{right})
is ${\cal J}=l$.   Each fluctuation in the series $\Delta=l+1$
saturates one of the bounds in (\ref{first})-(\ref{fourth}),
so these fluctuations should be dual to $1/4$ BPS operators.

\subsection{$AdS_3$ fluctuations inside $AdS_5$}
\label{wierd}

Let us now compute the conformal dimensions of the operators dual
to the scalars which describe the fluctuations of the probe inside
of $AdS_5$. From eqns.~\reef{quadBI} and~\reef{quadWZ} the action
for $X^2$ and $X^3$ is \beqa S_{2,3} &=& -T_{D3}L^4\int
d^4\sigma\sqrt{\bar{g}_4}\left(\frac{1}{2u^2}\partial_a
X^2\partial^aX^2 +\frac{1}{2u^2}\partial_aX^3\partial^aX^3\right)
\nonumber
\\
&+& T_{D3}L^4\int d^4\sigma\left(\frac{1}{u^4}\partial_\xi X^2
\partial_u  X^3 - \frac{1}{u^4}\partial_u X^2 \partial_\xi  X^3\right)
\label{x2x3act}               \,. \eeqa Writing $\sqrt{2\pi}X^i = X^i_l
\exp(il\xi)$ for $i=2,3$ and doing the integral over $\xi$ gives
\beqa S_{2,3}&=&- T_{D3}L^4\int
d^3\sigma\sqrt{g_3}\left(1+\frac{1}{2u^2}(g_3^{ab}\prt_aX_{-l}^i\prt_bX^i_l
+ l^2 X^i_{-l}X^i_l)\right) \nonumber\\
&+& T_{D3}L^4\int d^3\sigma
\frac{1}{u^4}\left(ilX_l^3\prt_{u}X_{-l}^2 -
ilX^2_l\prt_{u}X_{-l}^3  \right).\eeqa where $g^3_{ab}$ is the
metric for the $AdS_3$ geometry
\begin{align}
ds^2=\frac{1}{u^2}\left(-dt^2+dx_1^2+du^2\right) \,.
\end{align}
The $X^2, X^3$ mixing in the Wess-Zumino term is diagonalized by
working with the field $w_l \equiv X^2_l + i X^3_l$, in terms of
which the action is
\begin{align}
S_w = - T_{D3}L^4 \int d^3 \sigma \sqrt{g_3}\frac{1}{2u^2}(g^{ab}_3
\partial_aw_l^* \partial_b w_l + l^2 w^*_l w_l) \nonumber \\
+  T_{D3}L^4 \int d^3 \sigma \frac{1}{2u^4} \partial_u(lw^*_l
w_l) \,.
\end{align}
The usual action for a scalar field in $AdS_3$ is obtained by
defining $\tilde w_l = w_l/u$, giving
\begin{align}
S_w = - T_{D3}L^4 \int d^3\sigma
\sqrt{g_3}\frac{1}{2}\left(g_3^{ab}\prt_a\tilde w^*_l\prt_b\tilde w_l + (l^2
-4l + 3) \tilde w^*_l \tilde
w_l\right) \label{reg}\\
+  T_{D3}L^4 (l-1) \int d^3 \sigma \frac{1}{2}\partial_u (\frac{1}{u^2}
\tilde w^*_l \tilde w_l) \label{surf} \,.
\end{align}
The surface term (\ref{surf}) does not effect the equations of
motion, but will be significant later when we compute correlation
functions of the dual operators. Inserting the spectrum $m^2
=l^2 - 4l+3$ into the standard formula $\Delta = d/2
\pm\sqrt{d^2/4+m^2}$ gives \beq \Delta = 1\pm |l-2|\,.\eeq   This
gives two series of dimensions, $\Delta = l-1$ and $\Delta = 3-l$,
which are possible in the ranges of $l$ for which $\Delta$ is
non-negative.  The entry in the AdS/CFT dictionary for the series
$\Delta = l-1$ holds several remarkable surprises which we will
encounter later.

\subsection{Gauge field fluctuations \label{gff}}

We finally turn to the fluctuations of the world volume gauge
field. It is convenient to rescale fields according to
$\hat{F}_{ab}=2\pi l_s^2F_{ab}$ so that the gauge field
fluctuations have the same normalization as the scalars in the
previous subsection. We have \beqa S_{gauge}&=&-T_{D3}L^4\int
d^4\sigma\sqrt{\bar{g}_4}\frac{1}{4} \hat{F}_{ab}\hat{F}^{ab} \,.
\nonumber \\
&=&-T_{D3}L^4 \int d^4\sigma\sqrt{\bar{g}_4}\frac{1}{4}
\left(\hat{F}_{\alpha\beta}\hat{F}^{\alpha\beta} +
  2\hat{F}_{\xi\alpha}\hat{F}^{\xi\alpha}\right) \,.
\label{gaugeact} \eeqa In order to decouple the $AdS_3$ components
of the gauge field from that on the $S^1$ it is convenient to work
in the gauge $A_{\xi}=0$. Expanding the rest of the components in
Fourier modes on the $S^1$ so that
$A_{\alpha}=\hat{A}_{\alpha}e^{il\xi}$ the action becomes \beq
S_{gauge}=-T_{D3}L^4 \int d^3\sigma\sqrt{\bar{g}_3}\frac{1}{4}
\left(\hat{F}_{\alpha\beta}\hat{F}^{\alpha\beta} +
2l^2\hat{A}_{\alpha}\hat{A}^{\alpha}\right)      \,.
\labell{procact} \eeq The equations of motion are easily found to
be \beq D^{\alpha}\hat{F}_{\alpha\beta} + l^2\hat{A}_{\beta} = 0
\label{procaeqn} \eeq which are just the Maxwell-Proca equations
for a vector field with $M^2=l^2$. Using the standard relation
$\Delta = d/2 \pm \sqrt{(d-2)^2/4 +M^2}$ relating the mass of a
vector field to the dimension of its dual operator we find the
spectrum \beq \Delta_{\pm} = 1\pm l \label{vecspec} \eeq which for
$l>1$ requires us to choose the positive branch.

\section{Correlators from strings on the probe-supergravity background}
\setcounter{equation}{0}

The rules for using classical supergravity in an AdS background to
compute CFT correlators have a natural generalization to defect
CFT's dual to AdS probe-supergravity backgrounds. The generating
function for correlators in the defect CFT is identified with the
classical action of the combined probe-supergravity system with
boundary conditions set by the sources. This approach was used to
compute correlators in the dCFT describing the D3-D5 system in
\cite{DFO}. Without worrying yet about what the dual operators
are, we will do the same for the D3-D3 system here. In this
section we will highlight some peculiar features of this defect
CFT. First it will be shown that the correlators of operators dual
to probe fluctuations are independent of the 't Hooft coupling, at
least in the limit that the 't Hooft coupling is large. Second,
the two-point function of operators dual to one set of
fluctuations discussed in section (3.3) will be shown to vanish.
Correlators involving both defect and bulk fields are presented in
appendix A.

\subsection{Independence of the correlators on the 't Hooft coupling}
\label{sec41}
As in refs.~\cite{Boonstra, Bianchi, DFO} it is useful to work with a Weyl
rescaled metric
\begin{align}g_{MN}=L^2\bar{g}_{MN}\end{align}
where $L^2 =\sqrt{g_sN}l_s^2$.
In terms of the rescaled metric, the supergravity action
(\ref{IIB}) becomes \beq \frac{L^8}{2\kappa^2}\int
d^{10}x\sqrt{-\bar{g}}\left(R-\frac{1}{2}e^{2\Phi}
(\partial\Phi)^2 +\cdots\right)\sim N^2\int d^{10}x\sqrt{-\bar{g}}
\left(R-\frac{1}{2}e^{2\Phi} (\partial\Phi)^2 +\cdots\right)
\labell{scaledact} \eeq As in the usual AdS/CFT correspondence
correlation functions of gauge invariant operators in the bulk of
\mbox{$\N=4$} SYM at large 't~Hooft coupling are calculated by
expanding this action around the $AdS_5\times S^5$ vacuum of type
IIB. Here the presence of the probe D3-brane will make additional
contributions both through its world volume fields but also
through the pull backs of the $AdS_5\times S^5$ fields. Terms
involving the pull backs are dual to couplings between the bulk of
the field theory and the codimension 2 defect.
After Weyl rescaling the metric as above, the D3-brane probe
action $S_{DBI}+S_{WZ}$ becomes, \beq -L^4T_{D3}\int
d^4{\sigma}\sqrt{\bar{g}}(1+ {\rm fluctuations})\sim N\int
d^4{\sigma}\sqrt{\bar{g}}(1+ {\rm fluctuations}) \,.
\labell{scaleBI} \eeq Notice that the dependence on the 't~Hooft
coupling $\lambda = g_sN$ has completely dropped out of the
normalization of the action! Generic correlation functions
involving $n$ fields $\psi$ living on the D3-brane probe and $m$
fields $\phi$ from the bulk of $AdS_5$ arise from \beqa
S_{DBI}&=&N\int
d^4\sigma\left((\partial\psi)^2+\phi^m\psi^n\right)
\nonumber \\
&=&\int d^4\sigma\left((\partial\psi^{\prime})^2+
\frac{1}{N^{n/2+m-1}}\psi^{\prime}{}^n\phi^{\prime}{}^m\right)
\labell{scales} \eeqa where $\psi^{\prime}=N^{1/2}\psi$ and
$\phi^{\prime}=N\phi$ are the canonically normalized probe and
$AdS_5$ fields respectively.  The $N$ dependence of correlators
which follows from (\ref{scales}) is consistent with what one
expects in the planar limit.  It is interesting that none of these
correlation functions has any dependence on $\lambda$,  at least
for large $\lambda$ where the $AdS$ probe-supergravity description
is valid.

\subsection{Correlators from probe fluctuations inside
AdS$_5$: a surprise}\label{vanish}

Let us now compute the correlation functions associated to the
fluctuations $w_l$ of the probe brane inside $AdS_5$. For a
classical solution of the equation of motion,  
the action given by the sum of (\ref{reg})
and (\ref{surf}) is given by the surface term
\begin{align}
S_{cl} = - T_{D3}L^4 \int d^3 \sigma \frac{1}{2}\partial_u\left[\frac{1}{u}
\tilde w_l^* \partial_u \tilde w_l - (l-1)\frac{1}{u^2} \tilde
w_l^* \tilde w_l \right] \,.\label{clact}
\end{align}
The first term in this expression is of the standard form obtained in
AdS/CFT, for instance in \cite{FreedmanMathur}.
The new feature here which does not appear in standard AdS
computations is the extra surface term with coefficient $(l-1)$.
This term has dramatic consequences.
To see this we compute the two-point function of the operator dual to $w_l$
following the procedure of \cite{FreedmanMathur}. 
We introduce an $AdS_3$ boundary at $u=\epsilon$ and
evaluate the action (\ref{clact}) for a solution of the form
\begin{align}
w_l(u,\vec k) = K^{(l)} (u,\vec k) \, w_l{}^b (\vec k)
\end{align}
in momentum space satisfying the boundary conditions
\begin{align}
\lim_{u\rightarrow \epsilon} K^{(l)}(u,\vec k) =1, \qquad
\lim_{u\rightarrow \infty} K^{(l)}(u, \vec k) = 0 \,.
\end{align}
The solution of the wave equation with these boundary conditions is
\begin{align}
K^{(l)}(u,\vec k) = \frac{u}{\epsilon} \frac{{\cal K}_{\nu}(u |\vec
k|)}{{\cal K}_{\nu}(\epsilon |\vec k|)} \,,
\end{align}
where $\nu = \Delta -1$ and ${\cal K}_\nu(x)$ is the modified
Bessel function which vanishes at $x\rightarrow \infty$. Note that
this coincides with the calculation of \cite{FreedmanMathur} where in
this case $d=2$. The two-point
function is given by
\begin{align}
  \langle {\cal O}(\vec k) {\cal O}(\vec k') \rangle &\equiv
  - \frac{\delta^2}{\delta w_l{}^b(\vec k) \delta w_l{}^b(\vec k')} S_{cl}
 \Big|_{w_l{}^b = 0}   \nonumber \\
  &=-\frac{1}{\epsilon} \delta(\vec k + \vec k') \lim_{u\rightarrow \epsilon}
   \left[ \partial_u K(u,\vec k) - (l-1) \frac{1}{u} K(u,\vec k) \right] 
 \,,\label{fe}
\end{align}
with $S_{cl}$ the Fourier transform of (\ref{clact}). 

The non-local part of the two-point function is obtained by expanding ${\cal
  K}_\nu$ in a power series for small argument, keeping only the term which
scales like $\varepsilon^{ 2(\Delta-2)}$. The more singular terms give rise to
local contact terms of the form $\square^2 \delta(x-y)$ and are dropped. The
non-local contribution to the two-point function is given by
\begin{align} 
  \langle {\cal O}(\vec k) {\cal O}(\vec k') \rangle &=
\delta(\vec k + \vec k') \lim_{u\rightarrow
\epsilon}\left[-\epsilon^{-1}(\epsilon k)^{-1}\partial_u
\left(\frac{2^{-2(\Delta
-1)}\frac{\Gamma(2-\Delta)}{\Gamma(\Delta)}
(ku)^{\Delta}}{(k\epsilon)^{1-\Delta}}\right)\right. \nonumber \\
 &\left. \qquad \qquad \qquad\qquad  \ +\,  (l-1)\,
\epsilon^{-2}(\epsilon
k)^{-1}
\frac{2^{-2(\Delta-1)}\frac{\Gamma(2-\Delta)}{\Gamma(\Delta)}
(ku)^{\Delta}}{(k\epsilon)^{1-\Delta}} \right] \,. \label{second2}
\end{align}
The first of the two terms coincides exactly with the standard 
AdS calculation of \cite{FreedmanMathur}, whereas the second term
is an additional feature due to the presence of the probe brane.
Remarkably, there is an exact cancellation between the first and
the second term in (\ref{second2}) for the series
$\Delta = l-1$.  Thus for these fluctuations the usual calculation
does {\it not} give a power law correlation function of the form
$1/x^{2\Delta}$.  When we obtain the operators dual to these
fluctuations, it will become clear that one should not find a
power law. In particular, the lowest mode in this series is
the operator which parameterizes the classical Higgs branch.

\section{The conformal field theory of the D3-D3\\ intersection}
\label{sec3}
\setcounter{equation}{0}

Thus far we have only studied the dCFT on the D3-D3 intersection
in terms of its holographic dual, without ever writing the action.
In this section we will construct the action describing $N$
D3-branes orthogonally intersecting $N'$ D$3^{\prime}$-branes over
two common dimensions. In the notation of \cite{SkenderisTaylor}
this system is known as $(1 \vert \rm D3 \bot D3^{\prime})$. In
the discussion of holography it was assumed that $N\rightarrow
\infty$ with $g_{YM}^2N$ and $N'$ fixed,  such that the open strings
with both endpoints on the D3$'$-brane decoupled. We will not make
this assumption in constructing the action.

The $\N=4$ SYM $SU(N)$ theory located on the D3-branes and the
$\N=4$ SYM $SU(N')$ theory located on the D$3^{\prime}$-branes
couple to a $(4,4)$ hypermultiplet at a two-dimensional impurity.
Although $(4,4)$ supersymmetry is preserved, it is convenient to
work with $(2,2)$ superspace\footnote{A more complicated
alternative would be to work in harmonic $(4,4)$ superspace}. The
world volume of both stacks of D3-branes can be viewed as two
$\N=2, d=4$ superspaces, intersecting over a two-dimensional
$(2,2)$ superspace.  One of the $\N=2,d=4$ superspaces is spanned
by
\begin{align} {\cal X} \sim (z^+,z^-, w, \bar w,
\theta_i^{\alpha},\bar\theta^{i}_{\dot \alpha}) \,,
\end{align} with $z^\pm = X^0\pm X^1$ and $w=X^2+iX^3$. The index
$\alpha$ is a spinor index with values $1,2$, while the index $i$
accounts for the $\N=2$ supersymmetry and has values $1,2$. The
other $\N=2,d=4$ superspace is spanned by \begin{align}{\cal X}'
\sim (z^+, z^-,y,\bar y, \Theta_i^{\alpha},
\bar\Theta^i_{\dot\alpha})\,,\end{align} where $y= X^4 + iX^5$ and
one makes the identification\footnote{We put brackets around the
indices 1 and 2, which label the two Grassmann coordinates, in
order to distinguish these indices from spinor indices $\a,
\dot\a=1,2$.}
\begin{align}
\theta^1_{(1)} = \Theta^1_{(1)} \equiv \theta^+ \,, \\
\theta^2_{(2)} = \Theta^2_{(2)} \equiv \bar\theta^- \,.
\end{align} This is not the unique choice.  For instance
one could have written $\theta^2_{(2)} = \Theta^2_{(2)} \equiv
\theta^-$ which is related to the first choice by mirror symmetry
\cite{Hori}. The intersection is the $(2,2), d=2$ superspace
spanned by \begin{align} {\cal X} \cap {\cal X}' \sim (z^+,z^-,
\theta^+, \theta^-, \bar\theta^+, \bar\theta^-)\,.\end{align} All
the degrees of freedom describing the D3-D3$'$ intersection can be
written in $(2,2)$ superspace. For instance the D3-D3 strings,
which are not restricted to the intersection, can be described by
$(2,2)$ superfields carrying extra (continuous) labels $w,\bar w$.
Similiarly superfields associated to the
D$3^{\prime}$-D$3^{\prime}$ strings carry the extra labels $y,\bar
y$. Fields associated to D3-D$3^{\prime}$ strings are localized on
the intersection and have no extra continuous labels.

Due to the breaking of four-dimensional supersymmetry by the
couplings to the degrees of freedom localized at the intersection,
it is convenient to write the action in a language in which the
unbroken $(2,2)$ symmetry is manifest. This leads to a somewhat
unusual form for the four-dimensional parts of the action.  One
way to obtain this action is somewhat akin to
deconstruction~\cite{Nima}. The basic idea is to start with a
conventional $(4,4)$ two-dimensional action in $(2,2)$ superspace,
add an extra continuous label $w,\bar w$ to all the fields, and
then try to add terms preserving $(4,4)$ supersymmetry such that
there is a (non-manifest) four-dimensional Lorentz invariance. A
four-dimensional Lorentz invariant theory which has a
two-dimensional $(4,4)$ supersymmetry must also have ${\cal N} =4$
supersymmetry in four dimensions.   The procedure of constructing
a supersymmetric D-dimensional theory using a lower dimensional
superspace has been employed in several contexts
\cite{ArkaniGregoire,EGK,Hebecker}. The reader wishing to skip
directly to the action of the D3-D3 intersection in $(2,2)$
superspace may proceed to section \ref{intersectaction}.

\subsection{Four-dimensional
actions in lower dimensional superspaces} \label{decomp}

The approach of building four-dimensional Lorentz invariance
starting with a conventional $(4,4)$ supersymmetric theory is an
indirect but effective way to obtain the ${\cal N} =4, d=4$ super
Yang-Mills action in a two-dimensional superspace.  There is also
a more direct approach which gives a $(2,2)$ superspace
representation for the part of the ${\cal N} =4, d=4$ action
containing only the ${\cal N} = 2, d=4$ vector multiplet. The
${\cal N} =2, d=4$ vector multiplet has a straightforward
decomposition under two-dimensional $(2,2)$ supersymmetry.  On the
other hand, there is no off-shell ${\cal N} =2, d=4$ formalism for
the hypermultiplet, unless one uses harmonic superspace. We
demonstrate the decomposition of the vector multiplet below.  This
provides a useful check of at least part of the action appearing
in section \ref{intersectaction}.

\subsubsection{Embedding $(2,2)$, $d=2$ in $\N=2$, $d=4$}

We begin by showing how to embed $(2,2)$, $d=2$ superspace into
$\N=2$, $d=4$ superspace. The $\N=2, d=4$ superspace is
parametrized by ($z^+, z^-, w, \bar w$, $\theta_{(i)}^\a$, $\bar
\theta^{(i)}_{\dot\a}$). For the embedding let us redefine these
coordinates as
\begin{align} \label{coordinates}
\theta^+ &\equiv \theta_{(1)}^1, \quad
\thetasl^+ \equiv \theta_{(2)}^1\,, \nonumber\\
\bar \theta^- &\equiv \theta_{(2)}^2,\quad \thetasl^- \equiv
\theta_{(1)}^2 \,.
\end{align}
In the absence of central charges, the $\N=2$, $d=4$ supersymmetry
algebra is
\begin{align}
  \{Q_{(i)\a}, \bar Q^{(j)}{}_{\dot\b}\} &= 2 \rho^\m_{\a\dot\b} P_\m
\d_{i}^j,
  \qquad i,j=1,2 \,, \nonumber\\
  \{Q_{(i)\a}, Q_{(j)\b}\}&=\{\bar Q^{(i)}{}_{\dot\a}, \bar
  Q^{(j)}{}_{\dot\b}\}=0 \,
\label{d4algebra}
\end{align}
with Pauli matrices $\rho^\m$ given by Eq.~(\ref{Pauli}).  We define
supersymmetry charges $Q_+ \equiv Q_{(1)1}$, $\bar Q_- \equiv Q_{(2)2}$,
$\Qsl_+ \equiv Q_{(2)1}$, and ${{\Qsl_-} \equiv Q_{(1)2}}$.  Following the
methods of refs.~\cite{hori2, Hellerman}, we introduce a superspace defect
at
\begin{align}
w = 0 \nonumber, \quad \thetasl^+ =\thetasl^-= 0 \,,
\end{align}
which implies that the generators $P_2$, $P_3$, $\Qsl_\pm $, and $\bar
\Qsl_\pm$  are
broken. The unbroken subalgebra of (\ref{d4algebra}) is generated
by $Q_\pm$ and $\bar Q_\pm$ and turns out to be the $(2,2)$, $d=2$
supersymmetry algebra given by
\begin{align}
\{Q_\pm, \bar Q_\pm \} = 2 (P_0 \pm P_1) \,.
\end{align}
Other anticommutators of the $Q$'s vanish due to the absence of
central charges.

\subsubsection{$\N=2$, $d=4$ Super Yang-Mills action in $(2,2)$,
$d=2$ language} \label{sec2.2} In order to derive the $\N=2$
Yang-Mills action in $(2,2)$ language, we decompose the
four-dimensional $\N=2$ abelian vector superfield $\Psi$ in terms
of a two-dimensional $(2,2)$ chiral superfield $\Phi$, a twisted
chiral superfield $\Sigma$, and a vector superfield $V$. In the
abelian case, the twisted chiral superfield (see e.g.\
Ref.~\cite{Hori,phase}) is related to the vector multiplet by
\begin{align}
\Sigma \equiv \bar D_+ D_- V
\end{align}
and satisfies $\bar D_+ \Sigma = D_- \Sigma=0$.  The $(2,2)$
vector and chiral superfields can be obtained by dimensional
reduction of their ${\cal N} =1, d=4$ counterparts.

In appendix \ref{newappendix} we show that the $\N=2$, $d=4$
vector supermultiplet $\Psi$ decomposes into
\begin{align} \label{decomposition}
  \Psi = -i \Sigma + \thetasl^+ \bar D_+ \left( \bar \Phi -
    \pr_{\bar w}  V \right) + \thetasl^- D_- \left( \Phi -
    \pr_w V \right) + \thetasl^+ \thetasl^- G  \,
\end{align}
where $\pr_w$ is the transverse derivative and $G$ an auxiliary
$(2,2)$ superfield. An interesting result of the decomposition is
that the auxiliary field $D$ of the twisted chiral superfield
$\Sigma$ is related to the component $D'$ and transverse
derivatives of the components $v'_2$ and $v'_3$ of the
four-dimensional vector superfield,
\begin{align}
D=\frac{1}{\sqrt{2}} \left(D'+f'_{32} \right) \,, \label{Dterm}
\end{align}
where $f'_{32}=\partial_3 v'_2 -\partial_2 v'_3$.
Note that in distinction to the conformal field theory dual to the
$(2\vert \rm D3\perp D5)$ intersection studied in \cite{DFO, EGK}
there are no transverse derivatives like $\pr_w \phi'$ in
the auxiliary fields $F$ of the (2,2) superfield $\Phi$.

With the above decomposition of $\Psi$, we can now write down the
$\N=2$, $d=4$ (abelian) Yang-Mills action in (2,2) language.
Substituting Eq.\ (\ref{decomposition}) with $G=\bar D_+ D_- (i
\Sigma^\dagger + ...)$ into the usual form of the YM action, we
find
\begin{align}
  &\frac{1}{4\pi}{\rm\,Im\,}\tau \int d^4x d^2\theta_{(1)} d^2\theta_{(2)}
  \, \frac{1}{2} \Psi^2 \\
  &=\frac{1}{4\pi}{\rm\,Im\,} \tau \int d^4x d^4\theta \, \left(
    \bar \Sigma \Sigma + \bar \Phi \Phi +  \pr_{\bar w} V \Phi -
    \bar \Phi \pr_w V -  \pr_{\bar w} V \pr_w V \right) ,\nonumber
\end{align}
with $d^4\theta= \frac{1}{4} d\theta^+ d\theta^- d\bar\theta^+
d\bar\theta^- $. From this one can easily deduce the corresponding
non-abelian Yang-Mills action for vanishing $\theta$ angle,
\begin{align} \label{vectoraction}
  S^{\rm nonab}_{\rm YM}=\frac{1}{g^2} \int d^4x d^4\theta {\rm\,tr} \left(
    \Sigma^{\dagger}\Sigma + (\partial_{\bar w} + \bar \Phi) e^{V}
    (\partial_w + \Phi) e^{- V} \right) \,.
\end{align}

\subsection{The D3-D3 action in $(2,2)$ superspace} \label{intersectaction}

We now present the full action for the $(4,4)$ supersymmetric
theory describing the intersecting stacks of D3-branes. The action
has the form
\begin{align}
S= S_{\rm D3} + S_{\rm D3^{\prime}} + S_{\rm D3-D3^{\prime}} \,.
\end{align}
For each stack of parallel D3-branes we have separate actions,
$S_{\rm D3}$ and $S_{\rm D3^{\prime}}$, each of which correspond
to an $\N=4, d=4$ SYM theory with gauge groups $SU(N)$ and
$SU(N')$, respectively. The term $S_{\rm D3-D3^{\prime}}$
describes the coupling of these theories to matter on the
two-dimensional intersection.

In $(2,2)$ superspace, the field content of $S_{\rm D3}$ is as
follows. First, there is a vector multiplet $V(z^\pm, \theta^\pm,
\bar \theta^\pm; w, \bar w)$ or, more precisely, a continuous set
of vector multiplets labeled by $w, \bar w$ which are functions on
the $(2,2)$ superspace spanned by $(z^\pm, \theta^\pm,\bar
\theta^\pm)$.  The label $w = X^2+ i X^3$ parameterizes the
directions of the D3 world volume transverse to the intersection,
while $z^\pm = X^0 \pm X^1$ parameterizes the remaining
directions.  Under gauge transformations $V$ transforms as
\begin{align}
e^V \rightarrow e^{-i\Lambda^{\dagger}} e^V e^{i\Lambda} \,, \qquad
e^{-V} \rightarrow e^{-i\Lambda} e^{-V} e^{i\Lambda^{\dagger}} \,,
\end{align}
where $\Lambda$ is a $(2,2)$ chiral superfield which also depends
on $w, \bar w$. From $V$ one can build a twisted chiral (or field
strength) multiplet  as
\begin{align}
\Sigma = \frac{1}{2} \{ \bar {\cal D}_+, {\cal D}_- \} \,,
\end{align}
where ${\cal D}_\pm=e^{-V} D_\pm e^V$, ${\cal \bar D}_\pm=e^V \bar
D_\pm e^{-V}$. Additionally one has a pair of adjoint chirals
$Q_1$ and $Q_2$, transforming as
\begin{align} Q_i \rightarrow
  e^{-i\Lambda}Q_ie^{i\Lambda} \,.
\end{align}
Finally there is a $(2,2)$ chiral field $\Phi$ which transforms
such that $\partial_{\bar w} + \Phi$ is a covariant derivative:
\begin{align}
\partial_{\bar w} + \Phi \rightarrow e^{-i\Lambda}(\partial_{\bar w} + \Phi)
e^{i\Lambda} \,.
\end{align}
The complex scalar which is the lowest component of $\Phi$ is
equivalent to the gauge connection $v_2 + i v_3$ of the
four-dimensional SYM theory described by $S_{\rm D3}$.  This
structure was also seen in the explicit decomposition of the
ambient $\N=2, d=4$ vector field $\Psi$ under $(2,2), d=2$
supersymmetry discussed in section \ref{decomp}, cf.\ Eq.\
(\ref{redefinition}).

The action of the second D3-brane (D$3^{\prime}$) is identical to
that of the first D3-brane with the replacements
\begin{align}
w \rightarrow y \,,\quad V \rightarrow {\cal V}\,,\quad \Sigma
\rightarrow \Omega\,,\quad Q_i \rightarrow S_i\,,\quad \Phi
\rightarrow \Upsilon\,,
\end{align}
and is invariant under gauge transformations $\Lambda^{\prime}$.

The fields corresponding to D3-D$3^{\prime}$ strings are the
chiral multiplets $B$ and $\tilde B$,  which are bifundamental and
anti-bifundamental respectively with respect to $SU(N) \times
SU(N')$ gauge transformations;
\begin{align}
B\rightarrow e^{-i\Lambda} B e^{i \Lambda^{\prime}} \,, \qquad
\tilde B \rightarrow e^{-i\Lambda^{\prime}} \tilde B e^{i\Lambda} \,.
\end{align}

Using a canonical normalization ($V \rightarrow g V$ etc.), the
components of the action are as follows:
\begin{align}\ \label{action1}
S_{\rm D3} = \,&\frac{1}{g^2} \int d^2z  d^2w d^4\theta {\rm\, tr}
\left(\Sigma^{\dagger}\Sigma + (\partial_{w} +  g\bar \Phi) e^{gV}
(\partial_{\bar w} + g\Phi) e^{-gV}
+ \sum_{i=1,2} e^{-gV} \bar Q_i e^{gV} Q_i \right) \nonumber\\
+ & \int d^2 z d^2w  d^2 \theta  \epsilon_{ij} {\rm\, tr\,} Q_i
[\partial_{\bar w} + g \Phi, Q_j] + c.c
\end{align}
\begin{align} \label{action2}
S_{\rm D3^{\prime}}=\,&\frac{1}{g^2}\int d^2z  d^2y
d^4\theta {\rm\,tr} \left( \Omega^{\dagger}\Omega +
(\partial_{y} + g \bar \Upsilon) e^{g{\cal V}} (\partial_{\bar y} +
g\Upsilon)
e^{-g{\cal V}}
+ \sum_{i=1,2} e^{-g{\cal V}} \bar S_i e^{g{\cal V}} S_i \right) \nonumber\\
+ & \int d^2 z d^2y  d^2 \theta \epsilon_{ij} {\rm
\,tr\,} S_i [\partial_{\bar y} + g\Upsilon, S_j] + c.c
\end{align}
\begin{align} \label{defectaction}
S_{\rm D3- D3^{\prime}} = &\int d^2z d^4 \theta {\rm\, tr}
\left(e^{-g \cal V}\bar B e^{gV} B + e^{-gV} \bar {\tilde B}
e^{g \cal V} \tilde B \right) \nonumber \\
+ & \frac{ig}{{2}} \int d^2z d^2\theta {\,\rm tr}\left(
  B \tilde B Q_1 - \tilde B B S_1\right) + c.c.
\end{align}
with $d^4\theta= \frac{1}{4} d\theta^+ d\theta^- d\bar\theta^+
d\bar\theta^-$ and $d^2\theta = \frac{1}{2} d \theta^+ d
\theta^-$.

Some comments about $S_{\rm D3}$ are in order.  We have already
presented part of this action, as the first two terms in the
$S_{\rm D3}$ are given by Eq.\ (\ref{vectoraction}). Upon
integrating out auxiliary fields, $S_{\rm D3}$ can be seen to
describe the ${\cal N} =4$ SYM theory. To illustrate how
four-dimensional Lorentz invariance arises, consider the
superpotential $\epsilon_{ij} {\rm\, tr\,} Q_i [\partial_{\bar w}
+ \Phi, Q_j]$.  Upon integrating out the F-terms of $Q_1$ and
$Q_2$, one gets kinetic terms in the $X^2, X^3$ directions which
are the four-dimensional Lorentz completion of the kinetic terms
in the $X^0, X^1$ directions arising from $e^{-V} \bar Q_i e^V
Q_i$.

The form of $S_{\rm D3-D3'}$ is dictated by gauge invariance and
$(4,4)$ supersymmetry.  The geometric interpretation of various
fields can be seen from this part of the action.  The vacuum
expectation values for the scalar components of $Q_1$ and $S_1$
give rise to mass terms for the fields $B$ and $\tilde B$
localized at the intersection. There are also ``twisted'' mass
terms for $B$ and $\tilde B$ which arise when the scalar
components of the twisted chiral fields $\Sigma$ and $\Omega$ (or
equivalently of $V$ and ${\cal V}$) get expectation values.  One
expects  $B$ and $\tilde B$ fields to become massive when the
D3-branes are separated from the D3$'$-branes in the $X^{6,7,8,9}$
directions transverse to both. Thus we associate the scalar
components of $(Q_1, \Sigma)$ or $(S_1, \Omega)$ with fluctuations
in $(X^6+ iX^7, X^8+iX^9)$.

Note that in $(2,2)$ superspace, $Q_2$ and $S_2$ are not directly
coupled to the fields $B$ and $\tilde B$, although derivative
couplings arise after integrating out the F-terms of $Q_1$ and
$S_1$.  The scalar component of $Q_2$ describes fluctuations of
the D3-branes in the $y =X^4+iX^5$ plane parallel to the
D3$'$-branes. Similiarly the scalar components of $S_2$ describe
fluctuations of the D3$'$-branes in the $w =X^2+i X^3$ plane
parallel to the D3-branes.  When the orthogonal branes intersect,
a Higgs branch opens up on which the scalar components of $B$ and
$\tilde B$ have vevs (classically). The vanishing of the F-terms
of the chiral fields $S^1$ and $Q^1$ gives
\begin{align}
\frac{\partial W}{\partial q_1} = \partial_{\bar w} q_2 - g\delta^2(w) b
\tilde b &=0 \nonumber \\
\frac{\partial W}{\partial s_1}= \partial_{\bar y} s_2 -
g\delta^2(y) \tilde b b &= 0\,. \label{holomeq}\end{align} Because
of the geometric identifications $q_2 \sim y/\alpha^{\prime}$ and
$s_2 \sim w/\alpha^{\prime}$, the solutions of these equations
give rise to holomorphic curves\footnote{The holomorphic curves on
the Higgs branch were obtained in discussions with Robert Helling
and will be discussed more elsewhere.} of the form $w y =
c\alpha^{\prime}$, where $2\pi i c= g b\tilde b = g \tilde b b$.

\subsection{R-symmetries}

Recall that the isometries of the AdS backround are $SL(2,R)
\times SL(2,R) \times U(1) \times SU(2)_L \times SU(2)_R \times
U(1)$.   The $SU(2)_L \times SU(2)_R$ component is an R-symmetry
which acts as rotations in the $6,7,8$ and $9$ directions
transverse to all the D3-branes. The first $U(1)$ R-symmetry acts
as a rotation in the $w$ (or $23$) plane, while the second $U(1)$
acts as a rotation in the $y$ (or $45$) plane.  In the near
horizon geometry, the probe Kaluza-Klein momentum on $S^1$ is a
contribution to $J_{45}$.  The charge $J_{23}$ generates a
rotation in $AdS_5$ directions orthogonal to the probe.

Below we summarize the R-charges and engineering dimensions of the
fields of the D3-D3 intersection.
\begin{table}[ht]
\begin{center}
\begin{tabular}{ccllccc}
(4,4)& (2,2) &  components  & $(j_L, j_R)$ & $J_{23}$ & $J_{45}$ &
$\Delta$ \\
\hline & & $\sigma, q_1$ &
$(\frac{1}{2},\frac{1}{2})$ &$0$&$0$& $1$ \\
Vector& $Q_1, \Sigma$ &  $\psi_{q_1}^+, \bar\lambda_{\sigma}^+$ &
$(0,\frac{1}{2})$ &$\frac{1}{2}$& $-\frac{1}{2}$& $\frac{3}{2}$ \\
& &  $\psi_{q_1}^-, \bar\lambda_{\sigma}^-$ &
$(\frac{1}{2},0)$ &$ \frac{1}{2}$&$-\frac{1}{2}$& $\frac{3}{2}$ \\
& & $v_0, v_1$ & $(0,0)$ &$0$&$0$& $1$ \\
\hline
  & & $\phi$ & $(0,0)$ &$-1$&$0$& $1$ \\
Hyper& $Q_2, \Phi$ & $ q_2$ & $(0,0)$&$0$&$1$ & $1$ \\
& & $\psi_\phi^+, \bar\psi_{q_2}^+$ & $(\frac{1}{2},0)$ &
$-\frac{1}{2}$ & $-\frac{1}{2}$ &
$\frac{3}{2}$ \\
& & $\psi_\phi^-, \bar\psi_{q_2}^-$ & $(0,\frac{1}{2})$ &
$-\frac{1}{2}$&$-\frac{1}{2}$&
$\frac{3}{2}$\\
\hline
&  & $b$ & $(0,0)$ & $-\frac{1}{2}$ & $\frac{1}{2}$ & 0 \\
Hyper& $B, \tilde B$ & $\tilde b$ & $(0,0)$ & $-\frac{1}{2}$ &
$\frac{1}{2}$ & $0$ \\
& & $\psi_b^+, \bar\psi_{\tilde b}^+$  & $(\frac{1}{2},0)$ & $0$ &
$0$ &
$\frac{1}{2}$ \\
& & $\psi_b^-, \bar\psi_{\tilde b}^-$ & $(0,\frac{1}{2})$ & $0$ &
$0$ &
$\frac{1}{2}$ \\
\hline & &$\omega, s_1$ &
$(\frac{1}{2},\frac{1}{2})$ & $0$ &$0$& $1$ \\
Vector& $S_1, \Omega$ &  $\psi_{s_1}^+, \bar\psi_{\omega}^+$ &
$(0,\frac{1}{2})$ & $\frac{1}{2}$ & $-\frac{1}{2}$ & $\frac{3}{2}$ \\
& &  $\psi_{s_1}^-, \bar\psi_{\omega}^-$ &
$(\frac{1}{2},0)$ & $\frac{1}{2}$ & $-\frac{1}{2}$ & $\frac{3}{2}$ \\
& & $\tilde v_0, \tilde v_1$ & $(0,0)$ & $0$ &$0$ & $1$ \\
\hline
  & & $\upsilon$ & $(0,0)$ & $0$ & $1$ & 1 \\
Hyper& $S_2, \Upsilon$ & $ s_2$ & $(0,0)$ & $-1$ & $0$ & $1$ \\
& & $\lambda_\upsilon^+, \bar\psi_{s_2}^+$ & $(\frac{1}{2},0)$ &
$\frac{1}{2}$ & $\frac{1}{2}$ &
$\frac{3}{2}$ \\
& & $\lambda_\upsilon^-, \bar\psi_{s_2}^-$ & $(0,\frac{1}{2})$ &
$\frac{1}{2}$ & $\frac{1}{2}$ &
$\frac{3}{2}$\\
\end{tabular}
\caption{Field content of the D3-D3 intersection.}\label{rsym}
\end{center}
\end{table}

The $U(1)$ symmetries generated by $J_{45}$ and $J_{23}$ are
manifest in $(2,2)$ superspace.  The $U(1)$ generated by $J_{45}$
has the following action:
\begin{align}
&\theta^{+} \rightarrow e^{i\alpha/2} \theta^{+}\,,
&&B\rightarrow e^{i \alpha/2} B\,, &&
Q_2 \rightarrow e^{i\alpha} Q_2 \,,&&&&\nonumber\\
&\theta^{-} \rightarrow e^{i\alpha/2} \theta^{-}\,,
&&\tilde B \rightarrow e^{i \alpha/2} \tilde B\,,  &&
\Upsilon \rightarrow  e^{+i\alpha} \Upsilon \,, \nonumber\\
&y\rightarrow e^{i\alpha} y\,,
\end{align}
with all remaining fields being singlets. The $U(1)$ generated by
$J_{23}$ acts as
\begin{align}
&\theta^{+} \rightarrow e^{-i\alpha/2} \theta^{+}\,,
&&B\rightarrow e^{-i \alpha/2} B\,, &&
S_2 \rightarrow e^{-i\alpha} S_2\,,&&&& \nonumber \\
&\theta^{-} \rightarrow e^{-i\alpha/2} \theta^{-}\,,
&&\tilde B \rightarrow e^{-i \alpha/2} \tilde B\,,  &&
\Phi\rightarrow e^{-i\alpha}\Phi\,, \nonumber \\
&w\rightarrow  e^{-i\alpha} w\,.
\end{align}
The reader may be surprised that these R-symmetries act on the coordinates 
$w$
and~$y$.\footnote{Upon toroidal compactification of $w$ and $y$ the $U(1)$
  R-symmetry generated by $J_{23} + J_{45}$ is enhanced to $SU(2)$. Note 
that
  the $(4,4)$ supersymmetry algebra admits an $SU(2)_L \times SU(2)_R \times
  SU(2)$ automorphism \cite{DiaconescuSeiberg} which in the compactified 
case
  is also realized as a symmetry.}  However in the language of
two-dimensional superspace, these are continuous labels rather than 
space-time
coordinates. Recall also that $J_{23}$ (or $J_{45}$) is an R-symmetry of the
${\cal N} = 4$ algebra associated with one stack of D3-branes, but a Lorentz
symmetry for the orthogonal stack.

\section{Fluctuation--operator dictionary}
\setcounter{equation}{0}

In this section we find the map between fluctuations on the probe
D3-brane and operators localized at the defect.  The single
particle states on the probe correspond to meson-like operators
with strings of adjoint fields sandwiched between pairs of defect
fields in the fundamental representation.

\subsection{Fluctuations inside AdS$_5$}

The fluctuations of the probe D3-brane wrapping $AdS_3$ inside
$AdS_5$ are characterized by $w_l$, which is the Fourier transform
of $w=X^2+ iX^3$ on $S^1$. The associated R-symmetry charges are
$J_{23} = -1$ and $J_{45} = l$, while there are no charges with
respect to $SU(2)_L \times SU(2)_R$.  Recall that the possible
series of dimensions for operators dual to these fluctuations are
$\Delta=l-1$ and $\Delta = 3-l$.

\subsubsection{The $\Delta = l-1$ series and the classical Higgs
branch} \label{higgsbranch}

We now focus on the series $\Delta=l-1$.  In section
\ref{vanish}, we found that the usual $AdS$ computation of the
two-point function for this series does not give a power law behaviour.
Let us nevertheless determine the corresponding operators. In the
free field limit, a gauge invariant scalar operator which is
localized on the defect and has $\Delta = l-1, J_{23} =-1$ and
$J_{45} = l$ with no $SU(2)_L \times SU(2)_R$ charges is
\begin{align}
{\cal B}^l \equiv \tilde b q_2^{l-1} b \,. \label{wierdops}
\end{align}
This operator has dimension $\Delta = {\cal J}$, which saturates
the bounds (\ref{third}, \ref{fourth}) due to the superconformal
algebra.  An inspection of the supersymmetry variations of the
fundamental fields of the defect CFT also suggests that ${\cal
B}^l$ is a chiral primary. However this conclusion is erroneous.
In fact, ${\cal B}^l$ is not even a quasi-primary conformal field
due to the presence of the dimensionless scalars $b,\tilde{b}$. - In
other examples for probe brane holography were
the branes intersect  over more than two dimensions (for
instance for the D3-D5 intersection),  similar operators are in fact
chiral primaries. Here however,  massless scalar fields in two dimensions
have strong infrared fluctuations and logarithmic correlation
functions. In a unitary two-dimensional CFT, it is generally mandatory to
take derivatives of massless scalars or construct vertex operators
from them in order to obtain operators associated with states in the
Hilbert space.\footnote{In our case, due to the fact that $b$ and
$\tilde b$ transform in the fundamental and anti-fundamental
representations, it is not clear how to build a gauge covariant
vertex operator with power law correlation functions.} It may
therefore seem remarkable that operators such as (\ref{wierdops})
appear at all in the AdS/CFT dictionary.  Note that even though the 
apparent dimension of ${\cal B}^l$ is greater than zero for $l>1$,  the 
two-point functions do not have a standard power law behaviour.  
This can be readily
seen in perturbation theory,  where the scalars $b$ and $\tilde b$ give rise
to logarithmic terms in the two-point functions for ${\cal B}^l$.    

There is nevertheless a very simple interpretation for the
fluctuation $w_1$, the lowest mode in the $w_l$ series,
in the AdS background.  Recall that the
classical Higgs branch is parameterized by the vacuum value of the
field ${\cal B}^1 = \tilde b b$ and corresponds to the holomorphic
curves $wy\sim\langle \tilde{b}b\rangle = c$ via
eqns.~\reef{holomeq}. Furthermore, as discussed in section 2, the
probe brane can be embedded in $AdS_5 \times S^5$ so as to sit on
a holomorphic curve of precisely this form.  Thus it is natural to
expect that these holomorphic embeddings correspond to the
classical fluctuations $w_1$ about the $c=0$ embedding.

To see this is more detail let us elaborate on the relation
between the fluctuations $\tilde w_{1}$ and the classical Higgs branch.
Scalar fields in $AdS_3$ have the following behavior near the
$u\rightarrow 0$ boundary of $AdS_3$:
\begin{align}
\phi \sim u^\Delta f(z^\pm) + u^{2-\Delta} g(z^\pm) \,.
\end{align}
As is standard in the AdS/CFT duality (with Lorentzian signature)
non-normalizable classical solutions are to be interpreted as
sources for the corresponding operators, while the normalizable
solutions can be interpreted as specifying a particular state in
the Hilbert space \cite{Balasubramanian,Klebanov}.  Only the VEV
interpretation seems to make sense for the fluctuations $\tilde
w_l$ since, as shown in section \ref{vanish}, the two-point
functions calculated in the usual way with source boundary
conditions vanish. Let us examine the $l=1$ fluctuation for which
$\Delta = l-1= 0$, and consider the solutions $\tilde w_1 = c$
where $c$ is a complex number. Naively one might conclude that
this amounts to choosing $<\tilde b b> \sim c$. However since
$\Delta =0$, this solution is not normalizable, although it sits
right at the border of normalizability\footnote{Note that such solutions
have as much right to be considered in Euclidean signature,  since
they are non-singular at the ``origin'' of $AdS$, $u = \infty$.}.
This is a reflection of the fact that the quantum mechanical vacuum must 
spread out over the entire classical Higgs branch,  since the latter is 
parameterized by dimensionless scalars whose correlators grow
logarithmically with distance\footnote{This is the same
  spreading which accounts for the ``Coleman-Mermin-Wagner'' theorem \cite{CMW}
  preventing spontaneously broken continuous symmetries in two dimensions.}. 

Despite the lack of normalizability of the fluctuations $w_1 = c$,
the identification $c \sim \langle \tilde b b \rangle$ makes sense
at the classical level. This follows from the fact that the
solution $\tilde w_{1} = c$ corresponds to a holomorphic
embedding. To see this it is convenient to recall the following
coordinate definitions (with $L^2=1$):
\begin{align}
r= 1/u, \qquad z^{\pm} = X^0 \pm X^1, \quad w = u \tilde w = X^2 +
i X^3, \quad y = x^4+i x^5,
\end{align}
and define $\vec v = X^{6,7,8,9}$, in terms of which the D3-brane
metric is
\begin{align}
ds^2 = (1+ \frac{1}{r^4} )^ {-1/2}(-dz^+ dz^- + dw d\bar w) + (1 +
\frac{1}{r^4})^{1/2}(dy d\bar y + d\vec v^2) \,.
\end{align}
In the simplest case, the embedding of the probe D3$'$-brane is
given by $w=0, \vec v =0$. On the probe, $y= r\exp(-i\xi)$ where
$\xi$ is defined in (\ref{sphere}). Therefore $\tilde w_{1} = c$
implies
\begin{align}
w = u \tilde w_{1}e^{i\xi} = \frac{c}{r e^{-i\xi}} = \frac{c}{y} \,.
\end{align}
The holomorphic curve $w y =c$ is precisely that which arises from
(\ref{holomeq}), provided that
\begin{align}
b = \begin{pmatrix} v \cr 0 \cr  \vdots\end{pmatrix}  \qquad
\tilde b = \begin{pmatrix} v & 0 & \cdots \end{pmatrix}
\end{align}
with $gv^2 = c/(2\pi i)$.  In this background, the probe D3$'$-brane
combines with one of the $N$ D3-branes to form a single D3 on the
curve $w y =c$. In this sense the AdS field $w_1$ parameterizes the possible
embeddings of the probe brane within $AdS_5$ and the dual operator
$\tilde{b}b$ parameterizes the classical Higgs branch of the CFT.

As was noted earlier the curve $w y =c$ does not break the
superconformal symmetries. To see this, it is convenient to represent
$AdS_5$ by the
hyperboloid,
\begin{align}\label{hyperbol}
{\cal X}_0^2 + {\cal X}_5^2 - {\cal X}_1^2 - {\cal X}_2^2 - {\cal
X}_3^2 - {\cal X}_4^2 = 1
\end{align}
where
\begin{align}
ds^2 = -d{\cal X}_0^2 -d{\cal X}_5^2 + d{\cal X}_1^2 +d {\cal
X}_2^2 +d {\cal X}_3^2 +d {\cal X}_4^2 \,.
\end{align}
The coordinates on the Poincar\'{e} patch, $t , \vec x =
x^{1,2,3}$ and $r$, are related to these by
\begin{align}
{\cal X}_5 &= \frac{1}{2r} \left(1+r^2(1 + \vec x^2 -
t^2)\right), \qquad {\cal X}_0 = rt\,, \qquad
{\cal X}_{1,2,3} = r x^{1,2,3} \,, \\
{\cal X}_4 &= \frac{1}{2r} \left(1-r^2(1 + \vec x^2 - t^2)\right) \,.
\end{align}
The embedding $w y =c$,  or $x^2 + i x^3 = \frac{c}{r e^{i\xi}}$
can then be written as
\begin{align}
{\cal X}_2+ i{\cal X}_3 = ce^{-i\xi}  \,.
\end{align}
which when combined with eqn.~\reef{hyperbol} gives,
\begin{align}
{\cal X}_0^2 + {\cal X}_5^2 - {\cal X}_1^2  - {\cal X}_4^2 = 1+|c|^2.
\end{align}
This is exactly the hyperboloid which defines an $AdS_3$ spacetime
with radius of curvature $1+|c|^2$.
Further, this embedding is manifestly invariant under the isometry $SO(2,2)
\times SU(2)_L \times SU(2)_R \times U(1)'$.  The $U(1)'$ factor
is precisely that which appears in the superconformal algebra as a
combination of rotations in the $23$ and $45$ planes generated by
$J_{23} + J_{45}$.  This $U(1)'$ factor phase rotates $w$ and
shifts $\xi$ such that $we^{-i\xi}$ is invariant.

Quantum mechanically we expect the vacuum to spread out over the
entire classical Higgs branch, since it is parameterized by massless
two-dimensional fields.       
This differs from the situation on the
Coulomb branch, on which the orthogonal branes are separated in
the $X^{6,7,8,9}$ directions by giving VEV's to {\it
four}-dimensional fields $q_1, \sigma, s_1$ and $\omega$. Note
that on the Higgs branch one also has non-zero four-dimensional
fields, of the form $q_2 = c/w, s_2 = c/y$, however since the
asymptotic values of the fields are independent of $c$ in all but
two of the four world-volume directions,  we expect that there is
no obstruction to the wavefunction spreading out as a function of
$c$.  This suggests that the AdS/CFT prescription for computing
correlators should be modified to sum over embeddings of
holomorphic curves parameterized by $c$. A natural conjecture is
that the map between the generating function for correlators in
the CFT and the probe-supergravity action should have the form
\begin{align}
\langle e^{-J \hat O} \rangle = \int {\cal D} c\, e^{-S_{cl}(\phi,
c)}
\end{align}
where, as usual,  the probe-supergravity fields $\phi$ have
boundary behaviour determined by the sources $J$.  Note that the
classical Higgs branch is non-compact, and it is unclear to us
what the measure ${\cal D} c$ should be.\footnote{We expect that
one contribution to the measure should arise from the fact that
the $AdS_3$ metric induced on the curve $wy =c$ has effective
curvature radius $\sqrt{1 + c^*c}$.}

We note that the operators ${\cal B}^l$ have been proposed as
duals of the light-cone open string vacuum for D3-branes in a
plane-wave background \cite{SkenderisTaylor}.  The Penrose limit
giving rise to this background isolates a sector with large
$J_{45}$ in the defect CFT.  The light-cone energy in the plane
wave background corresponds to $\Delta - J_{45}$. For the
operators ${\cal B}^l$,  this quantity is negative: $\Delta -
J_{45} = -1$. Moreover we have seen that these operators are not
really chiral primaries (or even conformal fields).  Thus it is
not clear that they should be dual to the light-cone open string
vacuum. In fact it is not clear what the open string vacuum is,
due to the quantum mechanical spreading over the classical Higgs
branch, which corresponds different embeddings in the plane-wave
(or AdS) background.

\subsubsection{Fluctuations inside $AdS_5$: The $\Delta = 3-l$
series}

Next let us consider the series $\Delta = 3-l$ with $l \leq 1$. A
gauge invariant scalar operator on the defect having $\Delta =
3-l, J_{23} =-1, J_{45} = l$ with no $SU(2)_L \times SU(2)_R$
charges is
\begin{align} \label{Gl}
{\cal G}^l \equiv D_- \tilde b {q_2^{\dagger}}^{1-l} D_+ b
+ D_+ \tilde b {q_2^{\dagger}}^{1-l} D_- b
\end{align}
with the gauge covariant derivatives $D_\pm \equiv D_0 \pm D_1$.
Note that the two separate terms are necessary for parity
invariance under $z^+ \leftrightarrow z^-$. The fluctuations modes
$w_l$ are scalars rather than pseudoscalars. These operators
satisfy the bounds (\ref{first}) - (\ref{fourth}) and will be
shown to be descendants.

\subsection{Fluctuations inside $S^5$}
\label{chiralprimaries}

The fluctuations of the probe $S^1$ embedding inside $S^5$ are
characterized by the mode $V^m_l$ where $m=6,7,8,9$. These
fluctuations are scalars in the $(\frac{1}{2},\frac{1}{2})$
representation of $SU(2)_L \times SU(2)_R$ and have $J_{23} = 0$
and $J_{45} = l$. The possible series of dimensions are $\Delta =
1 \pm l$.  We need only consider $l\ge 0$ since ${V^m_l}^*=
V^m_{-l}$. In this case the sensible series of dimensions is
$\Delta = 1+l$. The only gauge invariant defect operator
consistent with this is
\begin{align}
{\cal C}^{\mu l} \equiv \sigma^\m_{ij} \left( \epsilon_{ik}\bar
\Psi^+_k q_2^l \Psi^-_j + \epsilon_{jk}\bar \Psi^-_k q_2^l
\Psi^+_i \right) \qquad (\m=0,...,3) \label{gurk}
\end{align}
where $\Psi^+_i$ and $\Psi^-_i$ are $SU(2)_L$ and $SU(2)_R$
doublets respectively,  given by
\begin{align}
\Psi^+_i = \begin{pmatrix} \psi_b^+ \cr \bar \psi_{\tilde b}^+
\end{pmatrix}
\qquad \Psi^-_i = \begin{pmatrix} \psi_b^- \cr \bar \psi_{\tilde
b}^-
\end{pmatrix}\,.\label{Psi}\end{align}
The index $\mu$ is an $SO(4)$ index and should not be confused
with a spacetime Lorentz index.  Note that (\ref{gurk}) is
invariant under parity, which exchanges the $SU(2)_L$ index $i$
with the $SU(2)_R$ index $j$, as well as $+$ with $-$.  This
operator saturates the bound (\ref{fourth}),  and is actually
$1/4$ BPS. For $l=0$, the operator is a pure defect operator which
satisfies both the bounds (\ref{second}) and (\ref{fourth}) and
thus is $1/2$ BPS. This operator will be shown to satisfy a
non-renormalization theorem to order $g^2$ in section 7, in
accordance with the results of section 4.1. - The operators
(\ref{Gl}) are obtained as two supercharge descendants of
(\ref{gurk}).

\subsection{Gauge field fluctuations}

The gauge field fluctuations as derived in section \ref{gff} transform
trivially under $SU(2)_L \times SU(2)_R$ and have $J_{23}=0$ and $J_{45}=l$.
If we pick the positive branch, the dimension of this operator is
$\Delta=l+1$.  On the field theory side,
the operator at the bottom of the tower with the same quantum
numbers is the current associated with a global $U(1)_B$ under which the
defect fields transform,
\begin{align}
  {\cal J}^M_B \equiv \bar\Psi^\a_i \rho^M_{\a\b} \Psi_i^\b + i \bar b
  \overleftrightarrow{D}^M b + i \tilde b\overleftrightarrow{D}^M
\bar{\tilde
    b} \qquad(M=0,1) \, \label{gaugeoperator},
\end{align}
with Pauli matrices $\rho^M$ defined by Eq.~(\ref{Pauli}), $\Psi$
as in (\ref{Psi}), and $\alpha, \beta\in \left\{+,-\right\}$.
Although this current is conserved and satisfies the BPS bound of the
superconformal algebra, it is not a quasi-primary of the $SO(2,2)$
global conformal symmetry. This is essentially due to the fact
that it is in the same (short) supersymmetry multiplet as the
dimensionless field $\bar b b + \tilde b \bar{\tilde b} $.

The contributions to (\ref{gaugeoperator}) involving $b$, $\tilde
b$ lead to logarithms in the correlation functions. These are
actually present even in the purely two-dimensional free field
theory obtained by setting $g=0$ and thus decoupling the 2d from
the 4d theory. In this case we have a bosonic current contribution
of the form
\begin{align}
  J_M^{\rm 2d} = i \bar b \partial_M b
  - i (\partial_M \bar b) b \,, \label{bosonicgaugeop}
\end{align}
which is conserved. For Euclidean signature, this current has a
correlator of the form
\begin{align}
  \langle J_M^{\rm 2d} (x) J_N^{\rm 2d} (0) \rangle \propto
  {\textstyle\frac{1}{2}} \ln (x^2\m^2) \frac{I_{MN}(x)}{x^2} + \frac{x_M
    x_N}{x^2} \,,\quad I_{MN}(x) = \delta_{MN} - 2 \frac{x_Mx_N}{x^2}\,,
\end{align}
where $I_{MN}(x)$ is the inversion tensor. (\ref{bosonicgaugeop})
satisfies $\partial^x_M \langle J_M^{\rm 2d} (x) J_N^{\rm 2d} (0)
\rangle =0$ for $x \neq 0$. Note that in complex coordinates we
have $\partial_{\bar z} J_z^{\rm
  2d} + \partial_{z} J_{\bar z}^{\rm 2d} =0$, where only the sum vanishes,
not each term separately, such that there is no holomorphic -
antiholomorphic
splitting.

On the supergravity side, it is not quite clear if the
current-current correlator obtained from the gauge field
fluctuations in section 3.4 is well-defined. In $AdS_3$, the
equation of motion for the gauge field leads formally to a
logarithmic propagator. This however does not satisfy the required
boundary condition to be identified as a bulk to boundary
propagator. A better understanding of the
role played by two-dimensional scalars in this model will be left
for future work.

\subsection{Summary and discussion of the AdS/CFT dictionary}

Table \ref{table2} summarizes the fluctuations of the KK modes and
their dual operators.\footnote{The conformal dimensions of the
dual operators are lowered by one in comparison with the
corresponding series in the D3-D5 system studied in \cite{DFO}.
This is simply because the operators are bilinears of defect
fundamental fields, whose conformal dimensions are lowered by 1/2
in comparison with corresponding defect fields in the D3-D5 case.}
The angular fluctuations of the probe $S^1$ embedding inside $S^5$
are dual to $1/4$ BPS primaries ${\cal C}^{\m l}$. The $\Delta =
3-l$ fluctuations of the embedding of $AdS_3$ inside $AdS_5$ are
dual to ${\cal G}^l$ which are two-supercharge descendants of
these primaries. The $\Delta = l-1$ fluctuations of the embedding
of $AdS_3$ inside $AdS_5$ are not dual to conformal operators
which correspond to states in the Hilbert space. Naively the dual
operators ${\cal B}^l$ look like $1/2$ BPS (chiral) primaries, but
in fact they contain massless defect scalars which do not give
rise to power law correlation functions. These massless scalars
and their dual fluctuations include an entry ${\cal B}^1$ which
parameterizes the classical Higgs branch.  The fluctuations ${\cal
B}^l$ for $l>1$ correspond to other holomorphic curves $w =
d/y^{l-1}$, however we do not (as yet) have a clear interpretation
for these in the defect CFT. Lastly, the operator ${\cal J}^M_B$
which is dual to the gauge field fluctuations on $AdS_3$ is a
descendant of the dimensionless operator $\bar b b + \tilde b
\bar{\tilde b}$, which has a logarithmic two-point function and is
not a primary operator although formally it trivially satisfies
the BPS bounds.

\begin{table}[!h]
\begin{center}
\begin{tabular}{c|c|c|l|c|c}
fluctuations & $\Delta$ & $l$ & $(j_1,j_2)_{\cal J} $  & operator
&
interpretation\\
\hline
$S^1 \subset S^5$ & $l+1$ & $l \geq 0$ &
$(\frac{1}{2},\frac{1}{2})_l$
& ${\cal C}^{\m l}$ & $1/4$ BPS primary \\
$AdS_3 \subset AdS_5$ & $3-l$ & $l \leq 1$ &  $(0,0)_{l+1}$ &
${\cal G}^l$&
descendant \\
& $l-1$ & $l \geq 1$&$(0,0)_{l+1}$ &  ${\cal B}^l$ &  classical
Higgs branch\\
gauge field & $l+1$ & $l \geq 0$ &  $(0,0)_l$ & ${\cal
J}_B^{M l}$ &
---
\end{tabular}
\caption{Summary of fluctuation modes and field theory operators
with coincident quantum numbers.} \label{table2}
\end{center}
\end{table}

\section{Nonrenormalization theorem}
\setcounter{equation}{0}

In section \ref{sec41} we found from considering strings on the
probe-supergravity background that correlators of both probe and bulk fields
should be independent of the 't~Hooft coupling $\lambda=g_{YM}^2N$. In
general, the weak and strong coupling behaviour do not have to be related.
Nevertheless, the remarkable result of complete 't~Hooft coupling 
independence
of the correlators at strong coupling suggests that nonrenormalization
theorems may be present in the defect conformal field theory.  In this 
section
we study the nonrenormalization behaviour of the correlators at weak 
coupling.
By showing the absence of order $g_{YM}^2$ radiative corrections to some of
the correlators, we give some field-theoretical evidence for the existence 
of
nonrenormalization theorems.  In particular, we consider the two-point
function of the chiral primary operator ${\cal C}^{\mu l}$ which is the 
lowest
component of a short representation of the (4,4) supersymmetry algebra 
derived
in Sec.~\ref{superalg}.

\subsection{Nonrenormalization of the two-point function involving
${\cal C}^{\m l}$}

\begin{figure}[!ht]
\begin{center}
\includegraphics{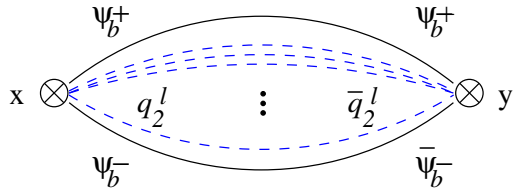}
\caption{One of the four graphs of the correlator $\langle {\cal
C}^l(x) \bar {\cal C}^l(y) \rangle$.}\label{rainbow}
\end{center}
\end{figure}

Let us consider the two-point correlator of the chiral primary ${\cal
C}^{\mu
  l}$. In the following we show that $\langle {\cal C}^{\mu l}
(x) \bar {\cal C}^{\mu l} (y) \rangle$ does not receive any corrections at
order $g_{YM}^2$ in perturbation theory.  It is sufficient to show this for
the component ${\cal C}^l \equiv {\cal C}^{1l}$ given by
\begin{align}
  {\cal C}^l &\equiv \psi^-_{\tilde b} q_2^l \bar
    {\psi}^+_{\tilde b} - \bar \psi^+_b q_2^l \psi^-_b + \bar \psi^-_{\tilde
      b} q_2^l {\psi}^+_{\tilde b} - \psi^+_b q_2^l \bar \psi^-_b
  \,.\label{C2}
\end{align}
The nonrenormalization of the other components is guaranteed by
the $SO(4)$ R-symmetry.  The tree-level graph of the two-point
function $\langle {\cal
  C}^l(x) \bar {\cal C}^l(y) \rangle$ is depicted in Fig.~\ref{rainbow}.
There are three other graphs contributing to this propagator
corresponding to the remaining three terms in Eq.~(\ref{C2}).

We show ${\cal O}(g^2)$ nonrenormalization for ${\cal C}^l$ with $l=0$ for
which $q_2$ exchanges are absent. The relevant propagators are
\begin{align}
&\langle v_M(x) v_N(y) \rangle = \frac{ \eta_{MN} }{ (2\pi)^2
(x-y)^2}, \qquad
\langle q_1(x) \bar q_1(y) \rangle = \frac{1 }{ (2\pi)^2 (x-y)^2}
\,,\label{bosonprop} \\
&\langle \psi_\a (x) \bar \psi_\b (y) \rangle =
\frac{i}{2\pi} \frac{\rho^M_{\a\b}  (x-y)_M}{(x-y)^2}
\,,
\end{align}
with $\eta_{MN}={\rm diag(+1,-1)}$, Pauli matrices $\rho^M
(M=0,1)$ defined in appendix~\ref{appA}, and defect coordinates $x, y$.
The four-dimensional propagators in Eq.~(\ref{bosonprop}) are
pinned to the defect. The Feynman rules for the vertices can be
read off from the defect action in component form derived in
appendix~\ref{appendixC}.

First we note that, similar as in $\N=4$, $d=4$ SYM theory \cite{Kovacs},
there are no one-loop self-energy corrections to the defect fermionic
propagator $\langle \bar \psi_b \psi_b \rangle$.  Self-energy corrections
involving a gaugino propagator are cancelled by those involving a
$\psi^{q_1}$
propagator which is the fermion of the superfield $Q_1$. There are also
self-energy graphs with $q_1$ and $\sigma$ propagators which arise from the
ambient scalars coupling to the defect. These cancel each other, too.

\begin{figure} [!ht]
\begin{center}
  \includegraphics{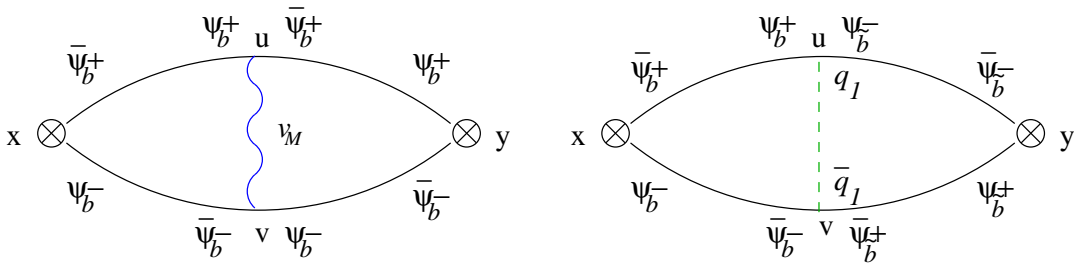}
\caption{First order corrections to the correlator $\langle  {\cal
C}^l(x) \bar {\cal C}^l(y) \rangle$ for $l=0$.}\label{rainbow2}
\end{center}
\end{figure}
However, we have two possible corrections from exchange graphs as shown in
Fig.~\ref{rainbow2}. Note that in Fig.~\ref{rainbow2}, two different
contributions to ${\cal C}^l$ ($l=0$) are depicted at the point $y$, which
originate from different terms in the sum (\ref{C2}).  These graphs include 
an
ambient gauge boson exchange and an ambient scalar exchange. There is no
$\sigma$ exchange contributing to the correlator $\langle {\cal C}^l(x) 
{\cal
  C}^l(y) \rangle$ (for $l=0$). In fact, it may be shown that for each of 
the
components of ${\cal C}^{\m l}$, there is either a $\sigma$ or a $q_1$
exchange. For all of the components, the vector exchange is cancelled by one
of these scalar exchanges while the other one vanishes.

For the gauge boson exchange in Fig.~\ref{rainbow2}a we find the
contribution
\begin{align}
  \frac{1}{2} \int d^2u d^2v \,& \frac{\rho_{++}\cdot(x-u)}{2\pi
    (x-u)^2} (-\frac{1}{2} g \rho^M_{++}) \frac{ \eta_{MN} }{ (2\pi)^2
    (u-v)^2} (-\frac{1}{2} g \rho^N_{--}) \frac{\rho_{++}\cdot(u-y)}{2\pi
    (u-y)^2}
  \nonumber\\
  &\times \frac{\rho_{--}\cdot(x-v)}{2\pi (x-v)^2}
  \frac{\rho_{--}\cdot(v-y)}{2\pi (v-y)^2} \,.
\end{align}
The overall factor $\frac{1}{2}$ comes from the definition $v_M=
\frac{1}{\sqrt{2}} v'_M$.

Let us now consider the contribution from the $q_1$
exchange in Fig.~\ref{rainbow2}b which is given by
\begin{align}
- \int d^2u d^2v \,& \frac{\rho_{++}\cdot(x-u)}{2\pi (x-u)^2}
(\frac{1}{2} ig) \frac{1 }{ (2\pi)^2 (u-v)^2} (-\frac{1}{2} ig)
\frac{\rho_{--}\cdot(u-y)}{2\pi (u-y)^2}
\nonumber\\
&\times \frac{\rho_{--}\cdot(x-v)}{2\pi (x-v)^2}
\frac{\rho_{++}\cdot(v-y)}{2\pi (v-y)^2}   \,. \label{C3}
\end{align}
Note that the operator at the external point $y$ in the graph of
Fig.~\ref{rainbow2}b is the conjugate of the first term in Eq.~(\ref{C2})
which leads to the minus sign in front of the integral in Eq.~(\ref{C3}).
In Fig.~\ref{rainbow2}a both external vertices have a minus sign, whereas
in Fig.~\ref{rainbow2}b the vertices have opposite signs.
Since $\eta_{MN} \rho^M_{++}\rho^N_{--}=2$, the vector exchange exactly
cancels the contribution from the scalar exchange.

Nonrenormalization of correlators of ${\cal C}^{\m l}$ with $l
\geq 1$ is more difficult to show.  As was shown for the operators
${\rm Tr\,}X^k$ in $\N=4$ super Yang-Mills theory \cite{Kovacs, D'Hoker},
there are no exchanges between the ambient propagators $\langle
q_2(x) \bar q_2(y) \rangle$ within the correlator $\langle {\cal
C}^l (x) \bar {\cal C}^l (y) \rangle$.  However, one could think
of a gauge boson exchange between a fermionic defect and a bosonic
ambient propagator.  If we do {\em not} work in Wess-Zumino gauge
then there is an additional interaction of the defect fermions
with a scalar $C$ which is the lowest component of the gauge
superfield $V$.  Keeping this in mind we expect that a CD exchange
\cite{Kovacs} cancels the above gauge boson exchange. This will be
shown elsewhere.

\subsection{Vanishing of odd correlators of the BPS primaries ${\cal C}^{\mu 
l}$}

Another property of the BPS primaries ${\cal C}^{\mu l}$ is the
vanishing of all $(2k+1)$-point functions ($k \in \mathbb{N}$).
Only even $n$-point functions may differ from zero. On the gravity
side this can be seen by studying once more the BI action of the
probe D3-brane.  Due to the expansion of the cosines of the
angular fluctuations $\theta, \phi, \rho$, and $\chi$ in the
determinant, the BI action contains only even powers of the
fluctuations, see Eq.\ (\ref{quadBI}).  This implies vanishing odd
couplings for the Kaluza-Klein modes which, via the AdS/CFT
correspondence, implies vanishing odd $n$-point functions on the
field-theory side. In the dual conformal field theory these
Kaluza-Klein modes correspond to the BPS primary operators ${\cal
C}^{\mu l}$. Again we restrict to the component ${\cal C}^l \equiv
{\cal C}^{1 l}$.

On the field theory side too,  one finds for instance that the
three-point function $\langle {\cal
  C}^{l_1} (x) {\cal C}^{l_2} (y) {\cal C}^{l_3} (z) \rangle $ is absent.
This is due to a global $U(1)$ symmetry of the action,
\begin{align}
B &\rightarrow e^{i\frac{\phi}{2}} B \,,\qquad
\tilde B \rightarrow e^{i\frac{\phi}{2}} \tilde B\,,\qquad
Q_1 \rightarrow e^{-i\phi} Q_1 \,,\qquad
Q_2 \rightarrow e^{i\phi} Q_2  \,,
\end{align}
with all other fields being singlets under this symmetry. If we
choose $\phi=\pi$ then \mbox{${\cal C}^l \rightarrow (e^{i\pi})^{l+1}
{\cal C}^l$} and the three-point function transforms as
\begin{align}
  \langle {\cal C}^{l_1} (x) {\cal C}^{l_2} (y) {\cal C}^{l_3} (z) \rangle
  \rightarrow (-1)^{l_1+l_2+l_3+1} \langle {\cal C}^{l_1} (x) {\cal C}^{l_2}
  (y) {\cal C}^{l_3} (z) \rangle \,.
\end{align}
Since $l_1+l_2+l_3$ must be even, $(-1)^{l_1+l_2+l_3+1}=-1$ and $\langle
{\cal
  C}^{l_1} (x) {\cal C}^{l_2} (y) {\cal C}^{l_3} (z) \rangle$ vanishes.
Though we have restricted the discussion on ${\cal C}^{1l}$, the statement
also holds for the other components. This is guaranteed by the fact that
${\cal C}^{\m l}$ transforms as a vector under the $SO(4)$ R-symmetry group.

\section{Conclusions and open questions}
\setcounter{equation}{0}

We have presented the action and some of the elementary properties
of a defect conformal field theory describing intersecting
D3-branes,  including some aspects of the AdS/CFT dictionary.
There remain many interesting open questions,  of which we
enumerate a few below.

The defect conformal field theory requires further field-theoretic
analysis. One of the stranger features of this theory is that it
contains massless two-dimensional scalars with (presumably)
exactly marginal gauge, Yukawa, and scalar potential couplings.
It is not at all obvious that one can construct a Hilbert space
corresponding to operators with power law correlation functions,
due to the logarithmic correlators of the two-dimensional scalars.
It would be very interesting if one could show this to all orders
in perturbation theory.

As a precursor to including gravity into the holographic map, it
would be interesting to study the energy-momentum tensor of the
defect conformal field theory in detail. We did not find any
evidence of an enhancement of the two-dimensional $SO(2,2)$ global
conformal symmetry to a full infinite-dimensional conformal
symmetry on the two-dimensional defect. A study of the
energy-momentum tensor would allow us to address this question
conclusively at least from the field-theoretic side. For example,
if an enhancement did indeed occur it should manifest itself in
the form of a two-dimensional energy-momentum tensor which is
holomorphically conserved.

Another question concerns the light-cone open string vacuum for
D3-branes in the Penrose limit of the probe-supergravity
background which we have considered.  The operator proposed in
\cite{SkenderisTaylor} to correspond to the open string light-cone
vacuum is not really a chiral primary and gives negative
light-cone energy.   This operator is precisely the one given in
(\ref{wierdops}),  and contains the dimensionless scalars which
parameterize the Higgs branch.  One might instead propose the
operator ${\cal C}^{\m l}$, with $P_- = \Delta -l = 0$ as the dual
of the light-cone vacuum, however this is only $1/4$ BPS and is in
a non-trivial representation of the unbroken $SU(2)_L \times
SU(2)_R$ R-symmetry.  Presumably the subtleties regarding the
light-cone vacuum are related to the quantum spreading over
holomorphic embeddings $wy =c$ corresponding to the classical
Higgs branch of the defect CFT. While the origin of this spreading
is clear from the point of view of the dual defect CFT, and from
the difficulties in finding localized supergravity solutions for
intersecting D3-branes \cite{marolfpeet,peet,Gomberoff}, they are
not so clear from the point of view of a probe D3-brane embedded
in the plane-wave or $AdS$ backgrounds.

Although there is presumably no fully localized supergravity
solution for intersecting D3-branes, it would be surprising if
there is no closed string string description,  in which both
stacks of D3-branes are replaced by geometry.  The problem of
finding a closed string description of the theory raises a closely
related question of how new degrees of freedom appear when $1/N$
(or $g_s$) corrections are taken into account in
probe-supergravity background which we have considered.  In
constructing the holographic dual of the defect CFT, we have fixed
the number $N^{\prime}$ of D3-branes in one stack, while taking
the number $N$ of D3-branes in the orthogonal stack to infinity.
In this limit, the degrees of freedom on one four-dimensional part
of the world volume of the defect become free. The remaining
coupled degrees of freedom live on a four-dimensional world volume
and a two-dimensional defect, which are the boundaries of $AdS_5$
and the embedded $AdS_3$ respectively. Because the defect degrees
of freedom are in the fundamental representation,  the genus
expansion of Feynman diagrams resembles an open string world-sheet
expansion. When $1/N$ corrections are taken into account, the
decoupled degrees of freedom must somehow reappear. The defect
degrees of freedom become bi-fundamental fields with respect to a
$SU(N) \times SU(N^{\prime})$ gauge group. The genus expansion for
Feynman diagrams of the theory can now be viewed as a closed
string world-sheet expansion, where a new branch of the target
space has opened up.\footnote{A similar although not directly
related picture has been discussed in \cite{OoguriVafa}.}

Finally, the string theory realization of the defect CFT leads one
to expect that it exhibits S-duality.  It would be very
interesting to find some field theoretic evidence for this.  In
particular one would need to find the S-duals of the fundamental
degrees of freedom localized at the intersection.

\bigskip

\vspace{1cm}

{\bf \large  Acknowledgements}

\vspace{1em}

We are grateful to Glenn Barnich, Oliver DeWolfe, Dan Freedman,
Ami Hanany, Robert Helling, Marc Henneaux, Andreas Karch, Neil
Lambert, Joe Minahan, Carlos Nu\~{n}ez, Volker Schomerus, David
Tong, Paul Townsend and Jan Troost for useful discussions. The
authors are particularly indebted to Robert Helling and to David
Tong for helpful comments. The research of J.E., Z.G.~and I.K.~is
funded by the DFG (Deutsche Forschungsgemeinschaft) within the
Emmy Noether programme, grant ER301/1-2. N.R.C.\ is supported by
the DOE under grant DF-FC02-94ER40818, the NSF under grant
PHY-0096515 and NSERC of Canada.

\newpage

\appendix

{\noindent \Large \bf Appendix}

\vspace{1cm}

\section{Kinematics of 2d/4d defect conformal field theories}
\setcounter{equation}{0}

\subsection{Conformal symmetry \label{confs}}

Here we present some basic implications of conformal symmetry in a
four-dimensional field theory with a two-dimensional defect.

Consider four-dimensional Euclidean space with a two-dimensional
defect. The coordinates are given by $v_\mu=(\vec{z}, \vec{x})$
where $v_\mu$ are the four-dimensional coordinates, $\vec{z}_M$
are the two defect coordinates and $\vec{x}_\alpha$ are the
coordinates perpendicular to the defect. The conformal
transformations which leave the defect invariant are given by
translations and rotations within the defect plane, rotations in
the plane perpendicular to the defect and by inversions $v_\mu
\rightarrow v_\mu/v^2$. The conformal group is given by
$SO(3,1)\times SO(2)$. Under these transformations we have for two
points $v$, $v'$
\begin{equation}
(v-v')^2 \rightarrow \frac{(v-v')^2}{\Omega(v)\Omega(v')} \, ,
\quad \vec{x}_\alpha \rightarrow \frac{\vec{x}_\alpha}{\Omega(v)}
\, \quad \vec{x}'_\alpha \rightarrow
\frac{\vec{x}'_\alpha}{\Omega(v')} \, .
\end{equation}
Hence there is a dimensionless coordinate invariant of the form
\begin{equation}
\xi = \frac{(v-v')^4}{(\vec{x}\cdot \vec{x}) (\vec{x}' \cdot
\vec{x}')} \,.
\end{equation}
Note that in the defect plane, we have only a global conformal
symmetry associated with the Virasoro generators $L_ {-1}$, $L_0$
and $L_1$. One may wonder if there is an accidental
two-dimensional local conformal symmetry giving rise to a Virasoro
algebra. This is however not the case: A necessary condition for
the existence of a Virasoro algebra is the existence of a
two-dimensional conserved local energy-momentum tensor. This
requirement is not satisfied in the situation considered here
since only the four-dimensional energy-momentum tensor of the
combined four-dimensional and two-dimensional action
contributions, given by
\begin{equation} \label{fourt}
T_{\mu \nu}(v) \, = \, T^{\rm bulk}{}_{\mu \nu }(v) \, + \, T^{\rm
def}{}_{MN} (\vec{z})\, \delta_{M(\mu} \delta_{\nu)N} \,
\delta^{(2)}(\vec{x})\, \, ,
\end{equation}
is conserved, $\partial_\mu T_{\mu \nu} = 0$. Note that this
energy-momentum tensor is in agreement with the $(2,2)$
supercurrent (\ref{twoj}) since from (\ref{fourt}) we obtain by
integration over $x$
\begin{gather}
{\cal T}_{MN}(z) = \int\! d^2x \, T^{\rm bulk}{}_{MN} (x,z) \, +
\, T^{\rm def}{}_{MN} (z) \, ,
\end{gather}
which is contained as a component in ${\cal J}_M$ given by
(\ref{twoj}). ${\cal T}_{MN}(z)$ satisfies $\partial^z{}_M {\cal
T}_{MN} (z) = 0$. Nevertheless it is not a local traceless
two-dimensional energy-momentum tensor.

For a quasi-primary scalar operator of dimension $\Delta$ close to
the defect we have a one-point function
\begin{equation} \label{onepoint}
\langle {\cal O} (v) \rangle \, = \, \frac{A_{\cal
O}}{(\vec{x}\cdot \vec{x})^{\Delta/2}} \, .
\end{equation}
Near to the defect we have a boundary operator expansion of the
bulk operators in terms of the defect operators, which reads
\begin{equation}
{\cal O}(v) \, = \, \sum\limits_{n} \frac{B_{{\cal O}, \hat {\cal O}_n
}}{ (\vec{x} \cdot \vec{x})^{(\Delta - \hat \Delta_n)/2}} \, \hat
{{\cal O}}_n(\vec{z}) \, .
\end{equation}
This gives rise to a bulk-defect correlator
\begin{equation} \label{twopoint}
\langle {{\cal O}}(v) \hat {{\cal O}}_n(\vec{z}') \rangle \, = \,
\frac{B_{{\cal O}, \hat {{\cal O}}_n }}{ (\vec{x} \cdot
\vec{x})^{(\Delta - \hat \Delta_n)/2} (v-v')^{2\hat \Delta_n}}\, ,
\quad  (v - v')_\mu = (\vec{x}, \vec{z}-\vec{z}').
\end{equation}
For two operators of dimension $\hat \Delta_n$ on the defect, this
expression reduces to
\begin{equation}
\langle \hat {{\cal O}}_n(\vec{z}) \hat {{\cal O}}_n(\vec{z}')
\rangle \, = \, \frac{B_{{\hat {{\cal O}}_n}, \hat {{\cal O}}_n
}}{ (\vec{z}-\vec{z}')^{2\hat \Delta_n}}\,
\end{equation}
as expected.

\subsection{SUGRA calculation of one-point functions and bulk-defect two
point functions}

We now compute the space-time dependence  of the bulk one-point
and the bulk-defect two-point function using holographic methods
and show that their structure agrees with the general results
obtained from conformal invariance in section \ref{confs}.  The
one-point function of the bulk operator ${{\cal O}}_\D$ is the
integral of the standard bulk-boundary propagator in $AdS_5$
\cite{FreedmanMathur} over the $AdS_3$ subspace. We find
\begin{align}
\langle {{\cal O}}_\D(\vec x, \vec z) \rangle &= \int \frac{dw
d\vec w^2}{w^3} \frac{\G(\D)}{\pi^2 \G(\D-2)}
\left(\frac{w}{w^2+\vec x^2 + (\vec w-\vec z)^2} \right)^{\D}
\nonumber\\
&= \frac{1}{\vert \vec x \vert^{\D}} \frac{\G(\frac{1}{2}\D+1)
\G(\frac{1}{2}\D-1)}{2\pi\,\G(\D-2) \, (\D-1)} \,,
\end{align}
which converges for $\D > 2$. The scaling behaviour $\vert \vec x
\vert^{-\D}$ has been expected from the structure of the one-point
function (\ref{onepoint}) on the CFT side. Note that the DBI
action (\ref{scales}) of the D3-brane probe determines the scale
dependence on $N$ of the correlation functions of defect and bulk
operators $\hat {{\cal O}}_{\hat \D}$ and ${{\cal O}}_\D$.

The two-point function $\langle {{\cal O}}_\D (\vec x, \vec z)
\hat {{\cal O}}_{\hat \D} (0) \rangle$ is the integral over the
product of the bulk-boundary propagators $K_{\D}\big(w,(\vec
x,\vec z),(\vec 0,\vec w)\big)$ and $K_{\hat \D}\big(w,(\vec
0,\vec w),(\vec 0,\vec 0)\big)$ ,
\begin{align}
  \langle {{\cal O}}_\D(\vec x, \vec z) \hat {{\cal O}}_{\hat \D} (\vec 0)
\rangle =
  \frac{1}{N^{1/2}} \frac{\G(\D)}{\pi^2 \G(\D-2)} \frac{\G(\hat \D)}{\pi
    \G(\hat \D-1)} J( \vec x, \vec z; \D, \hat \D)
\end{align}
with the integral
\begin{align}
  J( \vec x, \vec z; \D, \hat \D)&=\int \frac{dw d\vec w^2}{w^3}
  \left(\frac{w}{w^2+\vec x^2 + (\vec w-\vec z)^2} \right)^{\D}
  \left(\frac{w}{w^2 + \vec w^2} \right)^{\hat \D} \, \nonumber\\
  &= \frac{1}{(\vec x^2 + \vec z^2)^{\D}} \int dw' d\vec w'^2
     \, w'^{\hat \D-3}
     \left(\frac{w'}{w'^2+\vec x'^2 + (\vec w'-\vec z')^2} \right)^{\D}\,.
\end{align}
In the last line we made use of the inversion trick
\cite{FreedmanMathur} by defining
\begin{align}
(w',0,\vec w') = \frac{1}{w^2+\vec w^2}(w,0,\vec w)  ,\qquad (\vec
x',\vec z') =\frac{1}{\vec x^2 + \vec z^2}(\vec x,\vec z) \,.
\end{align}
As in \cite{DFO}, we rescale $\vec w' =\vec z' + \sqrt{\vec x'^2 +
w'^2} \vec v$ and $w'=\vert \vec x' \vert u$ and find
\begin{align}
J( \vec x, \vec z; \D, \hat \D)&= \frac{1}{(\vec x^2 + \vec
z^2)^{\hat \D} \vert \vec x \vert^{\D -\hat \D}} \int du
\frac{u^{\hat \D-3+\D}}{(1+u^2)^{\D-2}}
\int d\vec v^2 \frac{1}{(1+\vec v^2)^{\D}} \nonumber\\
&=\frac{1}{(\vec x^2 + \vec z^2)^{\hat \D} \vert \vec x \vert^{\D
-\hat \D}} \frac{\pi}{2(\D-1)} \frac{\G[\frac{1}{2}(\D-\hat \D)]
                            \G[\frac{1}{2}(\D+\hat \D)]}{\G[\D-2]} \,.
\end{align}
This converges for $\D > \hat \D$.  The scaling $ 1/( (\vec x^2 +
\vec z^2)^{\hat \D} \vert \vec x \vert^{\D -\hat \D} )$ agrees
with the behaviour of the two-point function fixed by conformal
invariance, cf.\ Eq.\ (\ref{twopoint}).

The defect-defect correlator $\langle \hat {{\cal O}}_{\hat \D}(
\vec z) \hat {{\cal O}}_{\hat
  \D} (\vec 0) \rangle$ for a defect operator $\hat {{\cal O}}_{\hat
  \D}$ is given by Eq.~(17) in \cite{FreedmanMathur} with
$d=2$ and is independent of $N$. Let us stress again that none of
the above correlators depends on $\lambda$ in the strong coupling
regime.

\section{Multiplets in $(2,2), d=2$ superspace}\label{appA}
\setcounter{equation}{0}

In order to fix the notation we briefly summarize the component
expansions of the superfields in $(2,2), d=2$ superspace which can
be found in \cite{Hori}, for instance. We use chiral coordinates
$y^0, y^1, \theta^\pm, \bar \theta^\pm$ which are related to the
superspace coordinates $x^0, x^1, \theta^\pm, \bar \theta^\pm$ by
\begin{align}
y^M&=x^M + i \theta^+ \rho^M_{11} \bar \theta^+ + i \theta^-
\rho^M_{22} \bar
\theta^- \nonumber\\
  &=x^M + i \theta^+  \bar \theta^+ + (-1)^M i \theta^- \bar \theta^- \,,
\qquad M=0,1 \,,
\end{align}
where we use the Pauli matrices
\begin{align}\label{Pauli}
\rho^0\equiv\s^0=\begin{pmatrix} 1 & 0\\ 0 & 1 \end{pmatrix} ,\,
\rho^1\equiv\s^3=\begin{pmatrix} 1 & 0\\ 0 & -1 \end{pmatrix},\,
\rho^2\equiv\s^2=\begin{pmatrix} 0 & -i\\ i & 0 \end{pmatrix} ,\,
\rho^3\equiv\s^1=\begin{pmatrix} 0 & 1\\ 1 & 0 \end{pmatrix} .
\end{align}
Expansions of (2,2) superfields can be obtained by dimensional
reduction of $\N=1, d=4$ superfields in the $2$ and $3$ direction
and defining $\theta^+\equiv\theta^1= \theta_2$ and $\theta^-
\equiv \theta^2= -\theta_1$. In this way we find the expansions of
the chiral and the vector multiplet in Wess-Zumino gauge,
\begin{align}
  \Phi(y,\theta^\pm) &= \phi + \sqrt{2} \theta^+ \psi_+ +
  \sqrt{2} \theta^- \psi_- -2 \theta^+ \theta^- F \\
  V(y,\theta^\pm,\bar\theta^\pm) &=\,\, \theta^- \bar \theta^- (v_0 - v_1)+
  \theta^+\bar\theta^+ (v_0+v_1)
  - \theta^- \bar\theta^+ \s - \theta^+\bar\theta^- \bar \s\nonumber\\
  &\quad+i \sqrt{2} \theta^- \theta^+ (\bar\theta^- \bar\lambda_-+\bar 
\theta^+
  \bar\lambda_+)+i \sqrt{2}
  \bar\theta^+ \bar \theta^- (\theta^- \lambda_-+ \theta^+ \lambda_+) 
\label{Vexp}\\
  &\quad+2\theta^-\theta^+\bar\theta^+\bar\theta^- (D -i \partial^M v_M) \,.
  \nonumber
\end{align}
The scalar $\s$ is complex and is defined in terms of the
components $v_1$ and $v_2$ of the dimensionally reduced
four-vector $v_\m$ by $\s \equiv v_3+iv_2$. For the (abelian)
twisted chiral superfield $ \Sigma(y,\theta^\pm,\bar\theta^\pm)
\equiv \bar D_+ D_- V(y,\theta^\pm,\bar\theta^\pm) $ we find the
expansion
\begin{align}
\Sigma(y,\theta^\pm,\bar\theta^\pm) = & \,\sigma + i\sqrt{2}
\theta^+ \bar \lambda_+ - i\sqrt{2} \bar \theta^- \lambda_- + 2 \theta^+
\bar\theta^-  ( D - i f_{01}) + 2i \bar \theta^- \theta^-
( \pr_0  - \pr_1) \sigma \nonumber\\
& - 2 \sqrt{2} \bar \theta^- \theta^- \theta^+ (\pr_0 - \pr_1)
\bar \lambda_+ \,.
\end{align}

\section{Decomposing the $\N=2$, $d=4$ vector multiplet under $(2,2)$,
  $d=2$ supersymmetry}  \label{newappendix}
\setcounter{equation}{0}

We start from the decomposition\footnote{IMPORTANT NOTE: In this section the
  $\N=2$, $d=4$ superspace is parametrized by ($x^0, ..., x^3$,
  $\theta_i^{\alpha},\bar\theta^{i}_{\dot \alpha}$) and the defect is placed
  at $x^1 = x^2=0$ in contrast with our convention in the rest of the text
  (defect at $x^2=x^3=0$).}  of the vector multiplet $\Psi$ under $\N=1$,
$d=4$ which is given by an expansion in $\theta_{(2)}$ \cite{Lykken},
\begin{align} \label{N=1expansion}
  \Psi(\tilde y,\theta_{(1)},\theta_{(2)})=
  \Phi'(\tilde y,\theta_{(1)}) + i \sqrt{2}
  \theta^\a_{(2)} W'_\a(\tilde y,\theta_{(1)}) + \theta_{(2)}
  \theta_{(2)} G'(\tilde y,\theta_{(1)})\,\,,
\end{align}
where $\Phi'$, $W'_\a$, and $G'$ are chiral, vector, and auxiliary
$\N=1$ multiplets, respectively. The superfield $\Psi$ is a
function of the coordinate $\tilde y^\m$ which is related to
$x^\m$ by
\begin{align}
  \tilde y^\m = x^\m &+ i \theta^+ \s^\m_{11} \bar \theta^+ + i \thetasl^-
  \s^\m_{21} \bar \theta^+ + i \theta^+ \s^\m_{12} \bar \thetasl^-
  +  i \thetasl^- \s^\m_{22} \bar \thetasl^- \nonumber\\
  &+ i \bar \theta^- \s^\m_{22}  \theta^- + i  \thetasl^+
  \s^\m_{12} \theta^- + i \bar \theta^- \s^\m_{21} \bar \thetasl^+
  +  i \thetasl^+ \s^\m_{11} \bar \thetasl^+ \,,
\end{align}
where $\s^0$ is the identity matrix and $\s^a$ ($a=1,2,3$) are the
Pauli matrices.

Our goal is to find an expression for $\Psi$ in terms of $(2,2)$,
$d=2$ multiplets,
\begin{align} \label{expansion}
\Psi \equiv \Psi \vert_{\thetasl=\bar\thetasl=0} + \thetasl^+
(\Dsl_+ \Psi ) \vert_{\thetasl=\bar\thetasl=0} + \thetasl^-
(\Dsl_- \Psi ) \vert_{\thetasl=\bar\thetasl=0} +  \thetasl^+
\thetasl^- (\Dsl_+ \Dsl_- \Psi) \vert_{\thetasl=\bar\thetasl=0}
\,,
\end{align}
with $\thetasl=(\thetasl^+,\thetasl^-)$. In order to find the
coefficients of this expansion, we substitute the component
expansions of $\Phi'$, $W'_\a$ and $G'$ in (\ref{N=1expansion})
and use the coordinates \mbox{($\tilde y$, $\theta^\pm$,
$\thetasl^\pm$, $\bar \theta^\pm$, $\bar \thetasl^\pm$)} as
defined in (\ref{coordinates}). We find\footnote{ conventions:
$(\psi_1, \psi_2)=(\psi_+,\psi_-)$; $\psi^+=\psi_-,
\psi^-=-\psi_+$}
\begin{align}
  \Psi =&\, \phi' + \sqrt{2} \theta^+ \psi'_+ + \sqrt{2} \thetasl^- \psi'_-
  - 2 \theta^+ \thetasl^- F' \nonumber\\
  &+ i \sqrt{2} \bar \theta^- \left( - i \lambda'_- + \theta^+ D' + \theta^+ 
(
    f'_{12} -i f'_{03} ) + \thetasl^- (f'_{02}-f'_{32} +i f'_{10} - i 
f'_{13})
  \right. \nonumber\\
  & \left. \qquad\qquad - 2 \theta^+ \thetasl^- (\pr_1 \bar \lambda'^+ + i
    \pr_2 \bar \lambda'^+ + \pr_0 \bar
    \lambda'^- - \pr_3 \bar \lambda'^-) \right) \nonumber\\
  &+ i \sqrt{2} \thetasl^+ \left( - i \lambda'_+- \thetasl^- D' - \theta^+
    (f'_{02} + f'_{32} - i f'_{10} - i f'_{13}) + \thetasl^-
    (f'_{21} + i f'_{03}) \right. \\
  &\left.\qquad\qquad - 2 \theta^+ \thetasl^- (\pr_1 \bar \lambda'^- - i 
\pr_2
    \bar \lambda'^- + \pr_0
    \bar \lambda'^+ + \pr_3 \bar \lambda'^+) \right) \nonumber\\
  &- 2 \thetasl^+ \bar \theta^- \left(F'^* - i \sqrt{2} \theta^+ (\pr_1 \bar
    \psi'^- - i \pr_2 \bar \psi'^- + \pr_0 \bar \psi'^+ + \bar \pr_3 
\psi'^+)
    + 2 \theta^+ \thetasl^- \square_4 \phi'^* \right) \,. \nonumber
\end{align}
Note that all fields are functions of $\tilde y$ and we have to
expand this expression such that all fields become functions of
the chiral coordinates $y^M=x^M + i \theta^+  \bar \theta^+ + (-1)^M i
\theta^- \bar \theta^- \, (M=0,3)$.  Evaluating $\Psi$, $\Dsl_+ \Psi$, and
$\Dsl_- \Psi$ at $\thetasl^+={\thetasl^-=0}$ we obtain
\begin{align}
  \Psi \vert_{\thetasl=0} &= - i  \Sigma \,,\nonumber\\
  \quad (\Dsl_+ \Psi ) \vert_{\thetasl=0} &= \bar D_+ \left( \bar \Phi -
    \partial_{\bar x} V \right)\,, \quad (\Dsl_- \Psi )
  \vert_{\thetasl=0} =  D_- \left( \Phi - \partial_{x}
    V \right) \,, \label{coeff}
\end{align}
where $\partial_x \equiv\partial_1+i\partial_2$ is the derivative transverse
to the defect.  Here we defined the (unprimed) components of the (2,2)
superfields $\Sigma$, $\Phi$, and $V$ in terms of the (primed) components of
the $\N=1$, $d=4$ superfield $\Phi'$ and $W'_\a$ by
\begin{align}
&\sigma \equiv i \phi', \quad \bar \lambda_+ \equiv \psi'_+, \quad \lambda_-
\equiv - \lambda'_-, \quad D \equiv \frac{1}{\sqrt{2}} ( D' + f'_{12}),
\quad
f_{03} \equiv \frac{1}{\sqrt{2}} f'_{03}   \,, \nonumber\\
&\phi{} \equiv \frac{1}{\sqrt{2}} (v'_1+iv'_2),\quad \bar \psi_+
\equiv \lambda'_+, \quad \psi_- \equiv \psi'_-,\quad F{} \equiv F'
\label{redefinition}
\end{align}
If we substitute the coefficients (\ref{coeff})  back into
the expansion (\ref{expansion}) of $\Psi$, we find the
decomposition (\ref{decomposition}).

The appearance of $f'_{12}$ in the definition (\ref{redefinition})
of the auxiliary field $D$ is required by $(2,0) \subset (2,2)$
supersymmetry. Consider the (2,0) supersymmetry transformation
rules for the spinor component $\lambda'^+$, the auxiliary field $D'$,
and the component $f'_{12}$ given by
\begin{align}
  \delta_{\eps}\lambda'^{+} &= i \eps^+ (D'+ f'_{12} -i f'_{03})\nonumber \\
  \delta_{\eps} D' &= \bar \eps^+ (\pr_0-\pr_3) \lambda'^+ - \eps^+
  (\pr_0-\pr_3) \bar \lambda'^+ - \bar\eps^+(\pr_1-i\pr_2)\lambda'^- - 
\eps^+
  (\pr_1-i\pr_2) \bar\lambda'^-
  \labell{susyvec} \nonumber\\
  \delta_{\eps} f'_{12} &= \eps^+ (\pr_1-i\pr_2) \bar\lambda'^- + \bar 
\eps^+
  (\pr_1-i\pr_2) \lambda'^- \,.
\end{align}
Of particular interest in Eq.~\reef{susyvec} are the non-standard
terms appearing in the variations of $\lambda'^{+}$ and $D'$
involving transverse derivatives, $\pr_2$ and $\pr_{1}$.  Note
that in dimensional reduction these terms would have simply been
set to zero.  The susy variation of $f'_{12}$ in $\delta_{\eps} D
\equiv \textstyle\frac{1}{\sqrt{2}}\delta_{\eps}(D'+ f'_{12})$
precisely cancels the non-standard terms in the variation of the
auxiliary field $D'$. This leads to the familiar (2,0) susy
variation for $D$,
\begin{align}
\delta_{\eps} D &= \bar \eps^+ (\pr_0-\pr_3) \lambda^+ - \eps^+
(\pr_0-\pr_3)
  \bar \lambda^+  \,.
\end{align}

\section{Impurity action in component form} \label{appendixC}
\setcounter{equation}{0}

In this appendix we derive the component expansion of the impurity
action in the decoupling limit which is given by
\begin{align} \label{impaction}
  S^{\rm dec}_{D3-D3^{\prime}} \equiv &\, S_{\rm kin} + S_{\rm superpot}
  \nonumber\\
  = &\int d^2z d^4 \theta \left(\bar B e^{gV} B + {\tilde B} e^{-gV} \bar
    {\tilde B} \right) +\, \frac{ig}{{2}} \int d^2z d^2\theta (\tilde B Q_1 
B)
  + c.c.
\end{align}
with $d^4\theta= \frac{1}{4} d\theta^+ d\theta^- d\bar\theta^+
d\bar\theta^-$ and $d^2\theta = \frac{1}{2} d \theta^+ d
\theta^-$.  Using the following expansions for the (2,2)
superfields $B$, $\tilde B$, $Q_1$,
\begin{align}
B &= b + \sqrt{2} \theta^+ \psi^b_+ +
  \sqrt{2} \theta^- \psi^b_- -2 \theta^+ \theta^- F^b \nonumber\\
\tilde B &= \tilde b + \sqrt{2} \theta^+ \psi^{\tilde b}_+ +
  \sqrt{2} \theta^- \psi^{\tilde b}_- -2 \theta^+ \theta^- F^{\tilde b}\\
Q_1 &= q_1 + \sqrt{2} \theta^+ \psi^{q_1}_+ +
  \sqrt{2} \theta^- \psi^{q_1}_- -2 \theta^+ \theta^- F^{q_1} \nonumber
\end{align}
as well as Eq.~(\ref{Vexp}) for $V$, the impurity action can be
expanded as
\begin{align}
  S_{\rm kin}&= \int d^2 z \big( \bar F^b F^b - \vert D_M b \vert^2 + i \bar
  \psi^b_- (D_0 + D_1) \psi^b_- + i\bar \psi^b_+ (D_0 - D_1) \psi^b_+
  \nonumber\\
  &\quad- \frac{g}{2} (\bar\psi^b_- \s \psi^b_+ + \bar \psi^b_+ \bar \s
  \psi^b_-) + \frac{ig}{{2}} (b \bar \psi^b_+ \bar \lambda_- - b \bar
  \lambda_+ \bar \psi^b_- - \bar b \lambda_- \psi^b_+ + \bar b \lambda_+
  \psi^b_- )
  \nonumber\\
  &\quad+ \frac{1}{2}(g D - \frac{1}{2} g_{YM}^2 \bar \sigma \sigma ) \bar b 
b
  \big)
  + (B \rightarrow \tilde B, g \rightarrow -g) \\
  S_{\rm superpot}&= \frac{ig}{{2}} \int d^2z \big( \tilde b F^{q_1} b +
  \tilde b \psi^{q_1}_- \psi^b_+ + {\psi}^{\tilde b}_+ \psi^{q_1}_- b +
  F^{\tilde b} q_1 b + \psi^{\tilde b}_- \psi^{q_1}_+ b +
  \psi_-^{\tilde b} q_1 \psi^b_+ \nonumber\\
  &\quad+ \tilde b q_1 F^b + \psi_+^{\tilde b} q_1 \psi_-^b + \tilde b
  \psi_+^{q_1} \psi_-^b \big) +c.c.
\end{align}
where we used the covariant derivative $D_M = \pr_M +
\textstyle\frac{i}{2} g v_M$ $(M=0,1)$.

For the ambient action we have the standard component expansion of
$\N=4, d=4$ SYM. Some of the components of the $\N=4$ ambient
vector field, which we gather in the (2,2) fields $V$ and $Q_1$,
couple to the impurity. The components of $V$ and $Q_1$ are
related to the components of the $\N=1, d=4$ superfields $V'$,
$\Phi'$, $\Phi'_1$, and $\Phi'_2$, which form the $\N=4$ vector
multiplet, by
\begin{align}
&\sigma \equiv i \phi', \quad \bar \lambda_+ \equiv \psi'_+, \quad \lambda_-
\equiv - \lambda'_-, \quad D \equiv \frac{1}{\sqrt{2}} ( D' + f'_{32}),
\quad
f_{01} \equiv \frac{1}{\sqrt{2}} f'_{01}   \,, \nonumber\\
&q_i \equiv \phi'_i, \quad \psi^{q_i}_\pm = \psi'^{\phi_i}_\pm,
\quad F^{q_i} \equiv F'^{\phi_i} \qquad(i=1,2) \,.
\end{align}

\section{Quantum conformal invariance}
\setcounter{equation}{0}

Here we give an argument that the action given by (\ref{action1}),
(\ref{action2}) and (\ref{defectaction}) does not receive quantum
corrections, such that it remains conformal to all orders in
perturbation theory. This argument is analogous to the discussion
of the 3d/4d case in \cite{EGK}, where more details on the
renormalization procedure may be found.

The argument for excluding possible quantum breakings of conformal
symmetry by defect operators relies on considering the (2,2)
supercurrent and its possible anomalies, and by making the
assumption that $(4,4)$ supersymmetry is preserved by the quantum
corrections. We begin by recalling the situation in $\N=1$, $d=4$
theories. In this case there is a supercurrent $J_{\dot \alpha
\beta} = \sigma^\mu{}_{\dot
  \alpha \beta} J_\mu$, which has the R current, the supersymmetry currents
and the energy-momentum tensor among its components. Potential
superconformal anomalies may be written in the form
\begin{equation}
\bar D^{\dot \alpha} J_{\dot \alpha \beta} \, = \, D_\beta S \, ,
\label{fourj}
\end{equation}
with $S$ a chiral superfield. When $S=0$, superconformal symmetry
is conserved.

By standard dimensional reduction to $(2,2)$ supersymmetry in two
dimensions, we obtain from (\ref{fourj}), as shown in \cite{West89},
\begin{equation}
(\gamma^M)_A{}^B \bar D_B {\cal J}_M \, = D_A \, {\cal S} \, ,
\end{equation}
where $M=\{1,2 \}$, $A,B=\{+,-\}$, $\gamma^M=\{\sigma^1, i
\sigma^2\}$ are the two-dimensional gamma matrices, ${\cal J}_M$
is the two-dimensional $(2,2)$ supercurrent and the possible
conformal anomaly is given by the $(2,2)$ chiral superfield ${\cal
S}$. ${\cal J}_M$ contains the 2d R-current, the four $(2,2)$
supersymmetry currents and the 2d energy-momentum tensor.

For 2d/4d models like the one given by (\ref{action1}),
(\ref{action2}) and (\ref{defectaction}), the classically
conserved two-dimensional supercurrent is given by
\begin{equation}
{\cal J}_M (z)\,=\, {\cal J}^{\rm def}{}_M (z) \, + \, \int\! d^2x
\; {\cal J}^{\rm bulk, 1}{}_{M} (x,z) \, + \, \int\! d^2y \; {\cal
J}^{\rm bulk, 2}{}_{M} (y,z)  \, .
\end{equation}
Let us first consider possible defect operator contributions to
the anomaly ${\cal S}$, which have to be gauge invariant and of
dimension 1. The possible defect contributions to the anomaly
${\cal S}$ are given by
\begin{gather} \label{twoj}
{\cal S}_D \, = \, {\rm Tr}\,   [ u\, \bar D^+ \bar D^- ( e^{-\cal
V}\bar B e^{V} B + e^{-V} \bar {\tilde B} e^{\cal V} \tilde B) \,
+ \, \, v \, (B \tilde B Q_1 - \tilde B B S_1) \, ] \, .
\end{gather}
It is important to note that there is no gauge anomaly term
contributing to this equation, since ${\rm Tr\,} \Sigma$ or $ {\rm
Tr\,} \Omega$, which would have the right dimension, are twisted
chiral and not chiral. $u$ and $v$ are coefficients which may be
calculated perturbatively. They are related to the $\beta$ and
$\gamma$ functions. From the standard supersymmetric
non-renormalization theorem we know that $v=0$ since the
corresponding operator is chiral. $u$ may be non-zero in a general
$(2,2)$ supersymmetric gauge theory. However $u$ and $v$ are
related by $(4,4)$ supersymmetry. Therefore if we assume that
$(4,4)$ supersymmetry is preserved upon quantization, $v=0$ also
implies $u=0$. Thus there are no defect contributions breaking
conformal symmetry.

We may also show that there are no  contributions from
four-dimensional operators to the conformal anomaly ${\cal S}$.
Such terms would have to originate from bulk action counter\-terms
which asymptotically fall off  with the distance from the
boundary. Asymptotically such anomaly contributions would be of
the form
\begin{equation}
{\cal S}_B \, \sim \, \int\!  d^2 w  \,|w|^{-s_1} \Lambda^{t_1}
{\cal O}_1 (w,z) \, + \, \int\!  d^2 y \,|y|^{-s_2} \Lambda^{t_2}
\, {\cal O}_2 (y,z)  \, ,
\end{equation}
with $\Lambda$ a regulator scale, and $s_i \geq 2$, $t_i\geq 0$
for $i=1,2$. However there are no such operators available in the
theory. From dimensional analysis, only ${\rm Tr\,} \Sigma$ or
${\rm Tr\,} \Omega$ would be possible for ${\cal O}_1$ or ${\cal
O}_2$, respectively, but again these are twisted chiral and not
chiral.
  Therefore we conclude that there a no
terms breaking $SO(2,2)$ conformal invariance, such that the
theory is conformal to all orders in perturbation theory.

\end{document}